\documentclass[preprint,aps,prd,showpacs,nofootinbib,floatfix,tightenlines]{revtex4}
\usepackage{graphicx}
\usepackage{amsmath}
\usepackage{amssymb}
\usepackage{bm}

\newcommand{\txtsm}[1]{\textrm{\scriptsize{#1}}}

\begin{document}

\title{\mbox{}\\[10pt]
Gluon fragmentation into quarkonium at next-to-leading order}

\author{Pierre  Artoisenet}
\affiliation{Centre for Cosmology, Particle Physics and Phenomenology (CP3), Universit\'e catholique de Louvain, Chemin du Cyclotron 2, B-1348 Louvain-la-Neuve, Belgium}
\author{Eric Braaten}
\affiliation{Department of Physics, The Ohio State University, Columbus,
Ohio 43210, USA}
 \begin{abstract}
We present the first calculation at next-to-leading  order (NLO) in $\alpha_s$
of a fragmentation function into quarkonium whose form at leading order is 
a nontrivial function of $z$, namely
the fragmentation function for a gluon into
a spin-singlet S-wave state at leading order in the relative velocity.
To calculate the real NLO corrections, we introduce
a new subtraction scheme that 
allows the phase-space integrals to be evaluated in 4 dimensions. 
We extract all ultraviolet and infrared divergences in the real NLO corrections
analytically by calculating the phase-space integrals 
of the subtraction terms in $4-2\epsilon$ dimensions.
We also extract the divergences in the virtual NLO corrections analytically,
and detail the cancellation of all divergences after renormalization. 
The NLO corrections have a dramatic effect on the shape 
of the fragmentation function, 
and they significantly increase the fragmentation probability. 
\end{abstract}
\date{\today}
\preprint{CP3-14-81}
\maketitle

\section{Introduction}

The production of a hadron in a high energy collision
is in general an extremely complicated problem
dominated by nonperturbative aspects of QCD.
There are several ways to simplify the problem 
in order to make a theoretical analysis more tractable.
One way is to consider inclusive production of the hadron,
summing over all possible additional hadrons in the final state.
Another simplification is to consider the production of the hadron 
with transverse momentum $p_T$ that is much larger 
than the momentum scale $\Lambda_{\rm QCD}$ of 
nonperturbative effects in QCD,
so that there are aspects of the problem that involve the 
small coupling constant $\alpha_s(p_T)$.
Another simplification is to consider a hadron whose constituents 
include a heavy quark whose mass $m$ is much larger 
than $\Lambda_{\rm QCD}$,
so that there are aspects of the problem that involve the 
small coupling constant $\alpha_s(m)$.
If the hadron is a heavy quarkonium, whose constituents are 
a heavy quark and antiquark, there are further simplifications
from the typical relative velocity $v$ 
of the constituents being small compared to 1.
The most theoretically tractable problem is the one in which 
all these simplifying features are combined:
the inclusive production of quarkonium at large $p_T$.

A rigorous factorization theorem for the inclusive production of a single hadron
at large $p_T$ was derived by Collins and Soper in 1981~\cite{Collins:1981uw}.
It states that in the inclusive cross section for producing a hadron $H$
at $p_T \gg \Lambda_{\rm QCD}$,
the leading power in the expansion in powers of $\Lambda_{\rm QCD}/p_T$ 
can be expressed as a sum of perturbative QCD (pQCD) cross sections 
for producing a parton convolved with {\it fragmentation functions}:
\begin{equation}
d \sigma[H + X] = \sum_i d\hat\sigma[i+X]\otimes D_{i\to H}(z).
\label{LP-frag}
\end{equation}
The sum extends over the types of partons (gluons, quarks, and antiquarks). 
The pQCD cross sections $d\hat\sigma$  are
essentially inclusive cross sections for producing the 
parton $i$, which can be expanded in powers of $\alpha_s(p_T)$,   
convolved with parton distributions if the colliding particles are hadrons. 
The nonperturbative factors $D_{i\to H}(z)$  are functions that give
the probability distribution for the longitudinal momentum fraction $z$
of the hadron $H$ relative to the parton $i$. 
The symbol ``$\otimes$'' in Eq.~\eqref{LP-frag} represents an integral over $z$.
Evolution equations for the fragmentation
functions can be used to sum large logarithms of $p_T/\Lambda_{\rm QCD}$ to all
orders in $\alpha_s$.

The factorization formula in Eq.~\eqref{LP-frag} applies equally well to 
heavy quarkonium with $m \gg \Lambda_{\rm QCD}$.
A proof of this  factorization theorem that deals 
with issues specific to quarkonium production
was first sketched by Nayak, Qiu, and Sterman in 2005 \cite{Nayak:2005rt}. 
It gives the leading power (LP) in the expansion in powers of 
$m/p_T$ and it applies only at $p_T \gg m$.
We will refer to this factorization theorem as the 
{\it LP factorization formula}.
In the case of cross sections summed over quarkonium spins,
the corrections are suppressed by a power of $m^2/p_T^2$
multiplied by logarithms of $p_T/m$.
The LP factorization formula has limited predictive power,
because the nonperturbative factors 
$D_{i\to H}(z)$ are functions of $z$ 
that must be determined from experiment.

In 1994, Bodwin, Braaten, and Lepage 
proposed the {\it NRQCD factorization formula},
which uses an effective field theory called nonrelativistic QCD
to separate momentum scales of order $m$ and larger 
from momentum scales of order $mv$ and smaller.
The theoretical status of the NRQCD factorization conjecture 
is discussed in Ref.~\cite{Bodwin:2013nua}.
The NRQCD factorization formula states that the inclusive cross section 
for producing a quarkonium state $H$ 
can be expressed as the sum of 
pQCD cross sections for producing a $Q\bar Q$ pair
with vanishing relative velocity multiplied by {\it NRQCD matrix elements}:
\begin{equation}
d \sigma[H + X] = \sum_n d \sigma[(Q \bar Q)_n + X]~
\langle  {\cal O}_n^{H} \rangle.
\label{NRQCD-fact}
\end{equation}
The sum extends over the color and angular-momentum channels of the $Q \bar Q$ pair. 
The pQCD cross sections $d \sigma$ are
essentially inclusive cross sections for producing the $Q \bar Q$ pair,
which can be expanded in powers of $\alpha_s(m)$,   
convolved with {\it parton distributions} if the colliding particles are hadrons. 
The nonperturbative factors $\langle{\cal O}_n^{H} \rangle$
are multiplicative constants that can be expressed as 
vacuum expectation values of 
four-fermion operators in nonrelativistic QCD \cite{Bodwin:1994jh}.
They scale as definite powers 
of the typical relative velocity $v$ of the $Q$ or $\bar Q$ in $H$. 
The NRQCD matrix element
$\langle{\cal O}_n^{H} \rangle$ is essentially the probability for a
$Q\bar Q$ pair created in the state $n$ to evolve into a final state 
that includes the quarkonium $H$.
Through heroic calculations over the past decade, the 
inclusive pQCD cross sections for producing a $Q\bar Q$ pair
in all the most phenomenologically relevant channels
have been calculated to next-to-leading order (NLO) in $\alpha_s$,
even for the most difficult case of hadron 
collisions~\cite{Butenschoen:2012px,Chao:2012iv,Gong:2012ug}.
In many cases, the NLO corrections are very large, 
which suggests that higher order corrections may also be important.
However the NLO calculations are sufficiently difficult that 
further improvement of the accuracy to next-to-next-to-leading order
seems to be out of the question. 

The predictive power of the LP factorization formula in Eq.~\eqref{LP-frag} 
can be increased by applying the 
NRQCD factorization conjecture to the fragmentation functions.
This  reduces the nonperturbative factors from functions of $z$ to 
multiplicative constants.
The fragmentation function for the parton $i$ to produce 
the quarkonium $H$ is expressed as a sum of 
functions of $z$ that can be calculated using pQCD
multiplied by NRQCD matrix elements:
\begin{equation}
D_{i\to H}(z) = \sum_n d_{i \to (Q \bar Q)_n}(z)~
\langle  {\cal O}_n^{H} \rangle.
\label{D-fact}
\end{equation}
The sum extends over the color and angular-momentum 
channels of a nonrelativistic $Q \bar Q$ pair. 
The pQCD functions $d_{i \to (Q \bar Q)_n}(z)$
can be expanded in powers of $\alpha_s(m)$.
The nonperturbative factors are
NRQCD matrix element $\langle{\cal O}_n^{H} \rangle$.
If the LP/NRQCD  factorization formula obtained by inserting 
Eq.~\eqref{D-fact} into Eq.~\eqref{LP-frag}
is expanded in powers of $\alpha_s(m)$, it
should reproduce the leading power in the expansion in powers of $m/p_T$
of  the NRQCD factorization formula in Eq.~\eqref{NRQCD-fact}.
The usefulness of the LP/NRQCD  factorization formula 
has proved to be limited at present collider energies.
Explicit calculations using the NRQCD factorization formula
have revealed that, in some channels, the LP cross section is
not the largest contribution until $p_T$ is almost an order of
magnitude larger than $m$~\cite{Chang:1996jt}.

An important recent development is the derivation of a factorization theorem
that extends the LP  factorization formula in Eq.~\eqref{LP-frag} 
to the next-to-leading power (NLP) of $m^2/p_T^2$. 
This factorization theorem was proven by 
Kang, Qiu, and Sterman \cite{Kang:2011zza,Kang:2011mg}.
A similar factorization formula has been derived by Fleming, Leibovich,
Mehen, and Rothstein using 
soft collinear effective theory \cite{Fleming:2012wy,Fleming:2013qu}.
In the {\it NLP  factorization formula},
the terms suppressed by $m^2/p_T^2$ are expressed as
a sum of pQCD cross sections for producing a collinear $Q \bar Q$ pair
convolved with {\it double-parton fragmentation functions},
which are nonperturbative probability
distributions in the longitudinal momentum fraction of the
quarkonium $H$ relative to the $Q \bar Q$ pair.
The predictive power of the NLP fragmentation formula 
can be dramatically increased by applying the NRQCD factorization formula 
to the double-parton fragmentation functions 
as well as the single-parton fragmentation functions  \cite{Ma:2013yla,Ma:2014eja}.
In the case of cross sections summed over quarkonium spins,
the corrections are suppressed by a power of $m^4/p_T^4$
multiplied by logarithms of $p_T/m$.
If the resulting NLP/NRQCD factorization formula 
is expanded in powers of $\alpha_s(m)$,
it should agree with the first few terms in the expansion 
in powers of $m/p_T$ of  the NRQCD factorization formula.
 
The NLP/NRQCD factorization formula 
opens the door to dramatic improvements in the accuracy of 
theoretical predictions for quarkonium production at 
very large $p_T$.  The factorization formula 
can be expressed as a triple expansion 
in powers of $\alpha_s$, $v$, and $m/p_T$.
NLP factorization incorporates subleading powers of $m/p_T$.
The NRQCD expansion includes subleading powers of $v$.
Accurate predictions also require including subleading powers of $\alpha_s$.
The aspects of the problem that are perturbative
are the pQCD cross sections for producing single partons, 
the pQCD cross sections for producing collinear $Q \bar Q$ pairs, 
the coefficient functions of NRQCD matrix elements
in the NRQCD expansions of both the 
single-parton fragmentation functions 
and the double-parton fragmentation functions,
and the evolution kernels for both sets of fragmentation functions.
It would be desirable to have all these ingredients calculated
to NLO in $\alpha_s$.

The NRQCD-expanded LP fragmentation formula was actually first applied to 
quarkonium production at large $p_T$ back in 1993, 
when the first fragmentation functions for S-wave quarkonium states
were calculated to leading order (LO) in $\alpha_s$ 
for channels that are leading order in $v$.
The fragmentation functions for a gluon into 
spin-singlet and spin-triplet S-wave states
at leading order in $\alpha_s$ and $v$
were calculated by Braaten and Yuan \cite{Braaten:1993rw,Braaten:1995cj}.
The fragmentation function for a heavy quark into 
spin-singlet and spin-triplet S-wave states at leading order in $\alpha_s$ and $v$
were calculated by Braaten, Cheung, and Yuan \cite{Braaten:1993mp}.
The fragmentation functions have since been calculated at LO in $\alpha_s$
for all the color and angular-momentum channels 
that are predicted by NRQCD factorization to be most phenomenologically relevant.
For the color-octet $^3S_1$ channel,
in which the LO fragmentation function is proportional to $\delta(1-z)$,
the fragmentation function has been calculated to 
next-to-leading order (NLO) in $\alpha_s$  \cite{Braaten:2000pc,Ma:2013yla}. 
In this paper, we present the first NLO calculation of a
fragmentation function whose form at LO is a nontrivial function of $z$:
the fragmentation function for a gluon into 
a spin-singlet S-wave state $\eta_Q$ at  leading order in $v$.

The outline of our paper is as follows.
In Section~\ref{sec:LO}, we calculate the LO fragmentation function
using Feynman rules introduced by Collins and Soper.
We also define some quantities that are useful in the NLO calculations.
In Section~\ref{sec:NLOreal}, we explain how the 
calculation of the real NLO corrections
can be simplified by introducing subtraction terms that cancel all the
ultraviolet and infrared divergences, 
allowing the phase-space integrals to be calculated in 4 dimensions.
We calculate the phase-space integrals of the subtraction terms 
analytically using dimensional regularization.
In Section~\ref{sec:NLOvirtual}, we describe the 
calculation of the virtual NLO corrections,
and we present analytic results for the 
poles that arise from the dimensionally regularized loop integrals.
In Section~\ref{sec:NLOrenorm}, we show how 
all the poles from phase space integrals and from loop integrals
are cancelled by renormalization 
of the parameters of QCD and renormalization of the 
operator whose matrix element defines the fragmentation function.
Some numerical illustrations of our results are
presented in Section~\ref{sec:numres}.
We discuss the prospects for the NLO calculation of all the other
phenomenologically relevant fragmentation functions  
in Section~\ref{sec:summary}.
Some details of the calculation of integrals 
at  NLO are presented in appendices.
In Appendix~\ref{sec:PhaseSpace}, we derive parametrizations of 
massless two-parton phase space integrals that are used
to integrate the subtractions terms for the real NLO corrections.
In Appendix~\ref{sec:LoopIntegrals}, we present the pole terms in 
the loop integrals with an eikonal propagator
that appear in the virtual NLO corrections.

\section{Leading-order fragmentation function}
\label{sec:LO}

In this section, we calculate the perturbative fragmentation function for 
$g \to Q \bar Q$, with the $Q \bar Q$ pair in a color-singlet $^1S_0$ state,
at leading order in $\alpha_s$.
We also introduce some related expressions that are 
useful in the calculation at next-to-leading order in $\alpha_s$.

\subsection{Feynman rules}

\begin{figure}
\center
\includegraphics[scale=0.4]{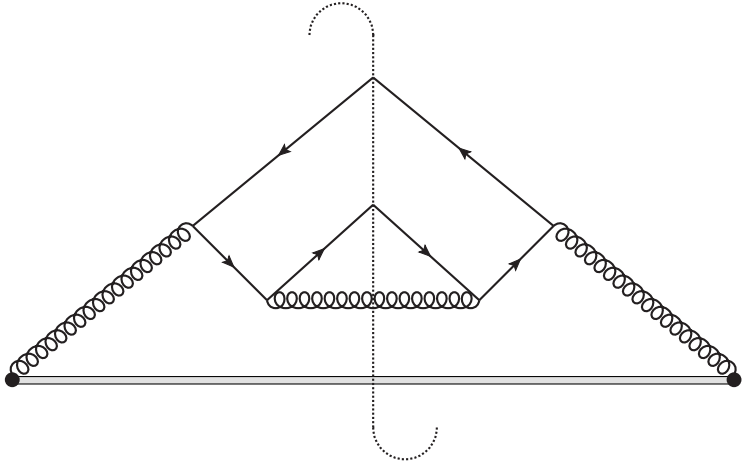}
\caption{
One of the 4 cut diagrams for gluon fragmentation into a 
color-singlet $^1S_0$ $Q \bar Q$ pair
at leading order in $\alpha_s$.
The eikonal line is represented by a double solid line. 
The dotted line is the cut.
The other 3 cut diagrams at leading order 
are obtained by interchanging the two gluon vertices on each side of the cut.}
\label{fig:cutdiagram}
\end{figure}

Gluon fragmentation functions can be calculated using  Feynman rules
derived by Collins and Soper in 1981~\cite{Collins:1981uw}.
The fragmentation function is expressed as the sum of all possible cut diagrams
of a particular form.  The diagrams include an eikonal line that
extends from a gluon-field-strength operator on the left side of the cut
to a gluon-field-strength operator on the right side.
Single virtual gluon lines are attached to the operators on the left side 
and the right side. The two virtual gluon lines from the operators are 
connected to each other by gluon and quark lines 
produced by QCD interactions, with 
possibly additional gluon lines attached to the eikonal line.
The cut passes through the eikonal line, the line for the particle
into which the gluon is fragmenting, and possibly additional gluon and quark lines.
An example of a cut diagram with the cut passing through the lines of
a heavy quark and antiquark 
and an additional gluon is shown in Figure~\ref{fig:cutdiagram}.

The Feynman rules for the cut diagrams are relatively simple \cite{Collins:1981uw}.
The 4-momentum $K$ of the gluon that is fragmenting
enters the diagram through the operator vertex on the left side of the eikonal line
and it exits through the operator on the right side.
Some of that momentum flows through the single virtual gluon attached to the operator
and the remainder flows through the eikonal line.
The operator at the left end of the eikonal line is labelled by a Lorentz index
$\mu$ and a color index $a$.
The operator at the right end of the eikonal line is labelled by a Lorentz index
$\nu$ and a color index $b$.  
The Feynman rules can be summarized as follows:
\begin{itemize}
\item
If the single virtual gluon line attached to the operator at the left end of the eikonal line
has momentum $q$, Lorentz index $\lambda$, and color index $c$,
the Feynman rule for the operator is
$-i(K.n g^{\mu \lambda} - q^\mu n^\lambda) \delta_{ac}$,
where $n$ is a light-like 4-vector.
\item
The attachment of an additional gluon line with 
Lorentz index $\beta$ and color index $c$ to an eikonal line
with color  indices $d$ and $e$ to the left and right
of the attachment
has the Feynman rule $g_s f^{cde}n^\beta$.
\item
The propagator for an eikonal line carrying momentum $q$
is $i/(q.n+ i \epsilon)$.
The Feynman rule  for a cut eikonal line carrying momentum $q$
is $2 \pi \delta(q.n)$.
\item
The remaining Feynman rules are those of QCD.
\end{itemize}

The particle into which the gluon is fragmenting has a specified 4-momentum.
In the case of fragmentation of a gluon into quarkonium, 
it is convenient to express that 4-momentum as $2p$.
The longitudinal momentum fraction $z$ of the quarkonium is 
\begin{equation}
\label{eq:defz}
 z =(2p).n/K.n.
\end{equation}
The fragmentation function is the sum of all cut diagrams 
contracted with $-g_{\mu\nu}$ and $\delta_{ab}$
and multiplied by the {\it Collins-Soper prefactor}~\cite{Collins:1981uw}
\begin{equation}
\label{eq:overalfac}
N_{\rm CS} = 
\frac{1}{(N_c^2-1)(2-2\epsilon)} \frac{z^{1-2\epsilon}}{2\pi K.n}.
\end{equation} 
The factors in the denominator include the number of color and spin states of 
a gluon in $D = 4-2 \epsilon$ dimensions.
The factor of $z^{D-3}$ arises from an integral over a transverse momentum.

\subsection{NRQCD Factorization}

The NRQCD factorization formalism \cite{Bodwin:1994jh} 
can be used to expand the 
fragmentation function for producing a quarkonium state
into a sum of matrix elements of NRQCD operators 
multiplied by perturbatively calculable coefficients. 
The NRQCD matrix elements scale as definite powers of the 
relative velocity $v$ of the heavy quark in the quarkonium.
For a $^1S_0$ quarkonium state  $\eta_Q$,
the matrix element that is leading order in $v$ was denoted 
$\langle{\cal O}_1(^1S_0)\rangle^{\eta_Q}$ in Ref.~\cite{Bodwin:1994jh}.
Within the vacuum-saturation approximation, it
can be interpreted as proportional to the square of the 
wavefunction at the origin for the quarkonium:
\begin{equation}
\label{eq:O1-R}
\langle{\cal O}_1(^1S_0)\rangle^{\eta_Q} = N_c |R(0)|^2/(2\pi).
\end{equation} 
The NRQCD factorization conjecture asserts that its coefficient,
which is a function of $z$,
can be calculated as a power series in $\alpha_s(m)$, 
where $m$ is the heavy quark mass.  
If we keep only the $\langle{\cal O}_1(^1S_0)\rangle^{\eta_Q}$ term,
the fragmentation function for $g \to\eta_Q$ can be expressed as
\begin{equation}
\label{eq:DNLOeta}
D_{g \to \eta_Q}(z) = 
\langle{\cal O}_1(^1S_0)\rangle^{\eta_Q} 
\left[ \alpha_s^2 d_{\rm LO}(z) + \alpha_s^3 d_{\rm NLO}(z) + \ldots \right].
\end{equation}
The function $d_{\rm LO}(z)$ in the leading-order term was calculated by
Braaten and Yuan in 1993 \cite{Braaten:1993rw}.
Our goal is to calculate the function $d_{\rm NLO}(z)$ 
in the next-to-leading order term.
This function also depends on renormalization and factorization scales 
that have been suppressed in Eq.~\eqref{eq:DNLOeta}.

The formation of the quarkonium $\eta_Q$
from the fragmentation of a gluon involves nonperturbative effects that
are represented by the sum of infinitely many Feynman diagrams.
Thus the fragmentation function $D_{g \to \eta_Q}(z)$
can not be calculated directly using perturbative QCD.
However the coefficient of $\langle{\cal O}_1(^1S_0)\rangle^{\eta_Q}$
in $D_{g \to \eta_Q}(z) $ can be determined from the perturbative
calculation of the fragmentation function for producing 
an appropriate $Q \bar Q$ state at a fixed order in $\alpha_s$.
The simplest choice is a $Q \bar Q$ pair in a color-singlet spin-singlet
state with zero relative momentum.
Its angular momentum quantum numbers are therefore $^1S_0$.
The perturbative fragmentation function $D_{g \to Q \bar Q}(z)$ 
for producing the $Q \bar Q$ pair
has the same form as in Eq.~\eqref{eq:DNLOeta}
but with a different prefactor:
\begin{equation}
\label{eq:DNLOQQ}
D_{g \to Q \bar Q}(z) = 
\langle{\cal O}_1(^1S_0)\rangle^{Q \bar Q} 
\left[ \alpha_s^2 d_{\rm LO}(z) + \alpha_s^3 d_{\rm NLO}(z) + \ldots \right].
\end{equation}
It can be calculated from cut diagrams for the gluon fragmentation function
in which the cut lines include $Q$ and $\bar Q$.
Given the normalization of the NRQCD operator ${\cal O}_1(^1S_0)$
defined in Ref.~\cite{Bodwin:1994jh}, 
the NRQCD matrix element for the $Q \bar Q$ pair is 
\begin{equation}
\label{eq:<O>QQbar}
\langle{\cal O}_1(^1S_0)\rangle^{Q \bar Q} = 2 N_c.
\end{equation}
If dimensional regularization is used to regularize ultraviolet and infrared 
divergences, this matrix element has no NLO corrections.
By dividing the perturbatively calculated fragmentation function 
$D_{g \to Q \bar Q}(z)$ by $2 N_c$, we can obtain the coefficient of
$\langle{\cal O}_1(^1S_0)\rangle^{\eta_Q}$ in Eq.~\eqref{eq:DNLOeta}.

The perturbative fragmentation function $D_{g \to Q \bar Q}(z)$
for producing a color-singlet spin-singlet $Q \bar Q$ pair with zero relative momentum
can be conveniently obtained by replacing the spinors from cuts through
the heavy quark and antiquark lines in a cut diagram by projection matrices.
The replacement rule for the  product of the spinors to the left of the cut is
\begin{equation}
\label{eq:project1S0}
v_j(p) \bar u_i(p) \longrightarrow 
\frac{1}{\sqrt{N_c}} \delta_{ij} \cdot 
\frac{1}{2\sqrt{2} m^{3/2}} (p\!\!\!/ - m) \gamma_5  (p\!\!\!/ + m).
\end{equation}
This projection matrix is the product of a color matrix with explicit indices $i$ and $j$
that projects the $Q \bar Q$ pair into a color-singlet state
and a Dirac matrix that projects it into a $^1S_0$ state.
The projection matrix $\Gamma_{ij}$ on the right side of 
Eq.~\eqref{eq:project1S0} satisfies 
Tr$(\Gamma_{ij} \gamma_0\Gamma_{ij}^\dagger \gamma_0) = 4m$,
which is the standard relativistic normalization for a particle of mass $2m$
in its rest frame.
With the projection matrix in Eq.~\eqref{eq:project1S0},
the Dirac structure on each side of the cut reduces to the trace
of Dirac matrices that include a single factor of $\gamma_5$.

We use dimensional regularization in $D=4-2\epsilon$ dimensions 
to regularize ultraviolet and infrared divergences.
Since the conventional definition of $\gamma_5$ is specific to 4 dimensions,
there is the possibility of an incompatibility between the definition 
of $\gamma_5$ and dimensional regularization.
One property of $\gamma_5$ that we will use 
is that the trace of a product of  $\gamma_5$ and fewer than four gamma matrices  
is 0.  In the LO and NLO diagrams for the fragmentation function,
this property can be used to reduce the Dirac trace 
in the amplitude on the left side of the cut to 
Tr$([ \gamma^\mu, \gamma^\lambda, \gamma^\rho ] p\!\!\!/ \gamma_5)$,
where $[ \gamma^\mu, \gamma^\lambda, \gamma^\rho ]$
is the antisymmetrized product of three gamma matrices 
whose 6 terms have coefficients $+1$ or $-1$.
The Dirac trace on the right side of the cut can similarly be reduced to 
Tr$([ \gamma^\nu, \gamma^\sigma, \gamma^\tau ] p\!\!\!/ \gamma_5)$.
After integrating over the momentum of the radiated gluon,
the only independent tensors that can be contracted with the product 
of these Dirac traces to give a scalar are 
$g_{\mu \nu} g_{\lambda\sigma} g_{\rho \tau}$
and $g_{\mu \nu} g_{\lambda\sigma} n_\rho n_\tau$.
In 4 dimensions, the Dirac trace from the left side of the cut is
\begin{equation}
\label{eq:Diractraceleft}
\mbox{$\frac16$} {\rm Tr} ([ \gamma^\mu, \gamma^\lambda, \gamma^\rho ] p\!\!\!/ \gamma_5)
= - i \epsilon^{\mu \lambda \rho \alpha} p_\alpha~{\rm Tr} (1).
\end{equation}
The Dirac trace on the right side of the cut gives a similar expression with 
Levi-Civita tensor $\epsilon^{\nu \sigma \tau \beta}$.
In 4 dimensions, the product of $ \epsilon^{\mu \lambda \rho \alpha}$
and  $\epsilon^{\nu \sigma \tau \beta}$ can be expressed as 
an antisymmetrized sum of products of four metric tensors with 24 terms.
With some of the more common prescriptions for $\gamma_5$,
these metric tensors can be interpreted as those for $D$ dimensions.
In this case, the contractions of the  two independent tensors
with the product of the two traces reduces to
\begin{subequations}
\begin{eqnarray}
\label{eq:gggTraces}
g_{\mu \nu} g_{\lambda\sigma} g_{\rho \tau} 
&\cdot &\mbox{$\frac16$}
{\rm Tr} ([ \gamma^\mu, \gamma^\lambda, \gamma^\rho ] p\!\!\!/ \gamma_5)
\cdot \mbox{$\frac16$}
{\rm Tr} ([ \gamma^\nu, \gamma^\sigma, \gamma^\tau ] p\!\!\!/ \gamma_5)
\nonumber\\
&&= (D-1)(D-2)(D-3) m^2 \big[ {\rm Tr} (1) \big]^2 ,
\\
\label{eq:ggnnTraces}
g_{\mu \nu} g_{\lambda\sigma} n_\rho n_\tau 
&\cdot& \mbox{$\frac16$}
{\rm Tr} ([ \gamma^\mu, \gamma^\lambda, \gamma^\rho ] p\!\!\!/ \gamma_5)
\cdot \mbox{$\frac16$}
{\rm Tr} ([ \gamma^\nu, \gamma^\sigma, \gamma^\tau ] p\!\!\!/ \gamma_5)
\nonumber\\
&&= - (D-2)(D-3) (p.n)^2 \big[ {\rm Tr} (1) \big]^2 .
\end{eqnarray}
\end{subequations}
The study of alternative prescriptions for $\gamma_5$ can ultimately be 
reduced to its effects on these two expressions.

\subsection{Born fragmentation function}

The fragmentation function for $g \to Q \bar Q$
can be calculated perturbatively from the cut diagrams 
in which the cut lines include $Q$ and $\bar Q$.
At leading order in $\alpha_s$, the cut diagrams are the diagram in Figure~\ref{fig:cutdiagram}
and three other diagrams.  One of the other diagrams is obtained 
by interchanging the vertices where the gluon from the operator
and the final-state gluon attach to the quark line on the left side of the cut.
The other two are obtained by making a similar interchange on the right side of the cut.
The cut lines are those for the $Q$ and $\bar Q$, 
the final-state gluon, and the eikonal line.
The final-state $Q$ and $\bar Q$ are on-shell with equal momenta $p$
and total longitudinal momentum fraction $z$.
The final-state gluon is on-shell with a momentum $q$
whose phase space must be integrated over.
The cut through the eikonal line gives a factor of $2 \pi \delta(K.n-(2p+q).n)$.

The amplitude corresponding to the sum of the two diagrams
on the left side of the cut can be written down using the Feynman rules:
\begin{eqnarray}
&& \frac{-ig_s^2}{(2p+q)^2[(p+q)^2-m^2]}  
\left[ K.n g_{\mu \lambda} - (2p+q)_\mu n_\lambda \right] \varepsilon^*_\beta(q)
\nonumber \\
&&  \hspace{2cm} \times
\bar u(p) \big[ (T^a T^c) _{ij} \gamma^\lambda (p\!\!\!/ + q\!\!\!/ - m) \gamma^\beta
- (T^c T^a) _{ij} \gamma^\beta (p\!\!\!/ + q\!\!\!/ + m) \gamma^\lambda \big] v(p) ,
\label{eq:ampI}
\end{eqnarray}
where $i$, $j$, and $c$ are the color indices of the final-state $Q$, $\bar Q$, 
and gluon.
After replacing the spinors by the projector in Eq.~\eqref{eq:project1S0}
and using the fact that the trace of the product of $\gamma_5$ 
and fewer than four gamma matrices is 0,
the amplitude can be reduced to
\begin{eqnarray}
\frac{-ig_s^2 \delta^{ac}}{2 (2 N_c)^{1/2} m^{1/2}(2p+q)^2p.q}  
\left[ K.n g_{\mu \lambda} - (2p+q)_\mu n_\lambda \right] \varepsilon^*_\beta(q) \; q_\delta
 \cdot  \mbox{$\frac16$}
{\rm Tr} \big[ [\gamma^\lambda, \gamma^\delta, \gamma^\beta] 
p\!\!\!/  \gamma_5   \big] .
\label{eq:ampII}
\end{eqnarray}
The cut diagram is obtained by multiplying this amplitude,
whose free indices are $\mu$ and $a$, by its complex conjugate 
with indices $\nu$ and $b$, integrating over the phase space of the gluon,
and summing over its color and spin states.
The fragmentation function $D_{g \to Q \bar Q}(z)$
is then obtained by contracting the cut diagram with
$\delta^{ab}(-g^{\mu \nu})$ and multiplying by the 
Collins-Soper prefactor in Eq.~\eqref{eq:overalfac}.

We denote the product of the differential phase space 
for the final-state gluon with momentum $q$
and the factor $2 \pi \delta(K.n-(2p+q).n)$ from the cut through the eikonal line
by $d\phi_{\rm Born}$.
We use dimensional regularization with $D = 4 - 2 \epsilon$ dimensions
to regularize both ultraviolet and infrared divergences. 
The integral over $q.n$ can be evaluated using the delta function 
from the cut through the eikonal line.
After integrating over the angles of the transverse components of $q$,
$d\phi_{\rm Born}$ reduces to a single differential:
\begin{equation}
 d\phi_{\rm Born} = 
 \frac{z^{-1+\epsilon} (1-z)^{-\epsilon}}{2 (4\pi)^{1-\epsilon} \Gamma(1-\epsilon) K.n}
 \left( s-\frac{4m^2}{z} \right)^{-\epsilon} ds,
 \label{eq:BornPS}
\end{equation}
where $s$ is the invariant mass of the $Q \bar Q g$ system:
\begin{eqnarray}
s = (2p+q)^2  .
\label{eq:szpq}
\end{eqnarray}
In Eq.~\eqref{eq:BornPS}, there is an implied Heavyside theta function
that imposes the constraint $s>4m^2/z$.

The fragmentation function for $g \to Q \bar Q$ at leading order in $\alpha_s$
can be expressed as
\begin{equation}
\label{eq:fragBorn}
D^{\rm (LO)}_{g \to Q \bar Q}(z) = N_{\rm CS}  \int d\phi_{\rm Born} 
 \mathcal{A}_{\rm Born}(p,q),
\end{equation}
where $N_{\rm CS} $ is the Collins-Soper prefactor in Eq.~\eqref{eq:overalfac} 
and the function $\mathcal{A}_{\rm Born}$ in the integrand is
\begin{eqnarray}
\mathcal{A}_{\rm Born}(p,q) &=&  
\frac{4 (1-2 \epsilon) (N_c^2-1) g_s^4 [(2p+q).n]^2}{N_c m s^2 (s-4m^2)^2}
\nonumber\\ 
&& \times
\left[ (1-2z+2z^2 -\epsilon)s^2   - 8 (z - \epsilon) m^2 s + 16(1-\epsilon) m^4 \right].
\label{eq:ABorn}
\end{eqnarray}
We will refer to this function as the {\it Born squared amplitude}.
The LO fragmentation function in $D$ dimensions is
\begin{eqnarray}
\label{eq:DLOeps}
D^{\rm (LO)}_{g \to Q \bar Q}(z)
&=& \frac{2(1-2 \epsilon)(4\pi)^\epsilon \alpha_s^2}{\Gamma(2-\epsilon) N_c m}
[z (1-z)]^{-\epsilon} 
\nonumber \\ 
&& \times 
\int_{4m^2/z}^\infty \!\!\!\!\!ds
\frac{(s-4m^2/z)^{-\epsilon}}{s^2}
\left[ 1 - \epsilon - 2 z(1-z) \frac{s (s-4m^2/z)} {(s-4m^2)^2} \right].
\end{eqnarray}

Since the integral in Eq.~\eqref{eq:DLOeps} has no divergences,
we can set $\epsilon =0$. The LO fragmentation function 
in 4 dimensions reduces to
\begin{equation}
\label{eq:dLOint}
D^{\rm (LO)}_{g \rightarrow Q\bar Q} (z) \Big|_{\epsilon = 0} = 
\frac{2 \alpha_s^2}{N_c m}
\int_{4m^2/z}^\infty \!\!\!\!\!ds
\frac{(1-2z+2z^2)s^2 - 8 z m^2 s + 16 m^4}{s^2 (s-4m^2)^2}.
\end{equation}
After evaluating the integral over $s$, the final result 
for the LO fragmentation function is
\begin{equation}
\label{eq:dLO}
D^{\rm (LO)}_{g \rightarrow Q\bar Q} (z) \Big|_{\epsilon = 0} = 
\frac{\alpha_s^2}{2 N_c m^3}
\left[2(1-z) \log(1-z) +3z-2z^2 \right].
\end{equation}
The NRQCD matrix element,
which is given by Eq.~\eqref{eq:<O>QQbar},
can be inserted by multiplying by 
$\langle{\cal O}_1(^1S_0)\rangle^{Q \bar Q}/(2 N_c)$.
Comparing with Eq.~\eqref{eq:DNLOQQ},
we can read off the function $d_{\rm LO}(z)$
in the fragmentation function for $g \to Q\bar Q$:
\begin{equation}
\label{eq:dLOcoeff}
d_{\rm LO}(z) = 
\frac{1}{4 N_c^2 m^3}
\left[2(1-z) \log(1-z) +3z-2z^2 \right].
\end{equation}
This same function $d_{\rm LO}(z)$ appears
in the fragmentation function for $g \to Q \bar Q$ in Eq.~\eqref{eq:DNLOeta}.
The leading-order fragmentation function calculated by Braaten and Yuan 
in 1993 \cite{Braaten:1993rw} can be reproduced by 
inserting the expression for $\langle{\cal O}_1(^1S_0)\rangle^{\eta_Q}$
in Eq.~\eqref{eq:O1-R}.

\subsection{Born squared amplitudes with uncontracted Lorentz indices}

To facilitate the calculation of the NLO corrections
to the fragmentation function, it is convenient to
generalize the  integration measure for the LO
fragmentation function in Eq.~\eqref{eq:DLOeps}
by allowing $q$ to be an arbitrary light-like vector.
The Collins-Soper prefactor in Eq.~\eqref{eq:overalfac} 
can be generalized to a function of $p$ and $q$:
\begin{equation}
\label{eq:overalfacBORN}
N_{\rm Born}(p,q) = 
\frac{1}{(N_c^2-1)(2-2\epsilon)}  \frac{1}{2\pi (2p+q).n)}   
\left( \frac{2p.n}{(2p+q).n} \right)^{1-2\epsilon}.
\end{equation}
The Born phase-space measure in Eq.~\eqref{eq:BornPS}
generalizes to
\begin{equation}
 d\phi_{\rm Born}(p,q) = 
 \frac{1}{2 (4\pi)^{1-\epsilon} \Gamma(1-\epsilon)}
\frac{(q.n)^{-\epsilon}}{(2p.n)^{1-\epsilon}} 
 \left( s-\frac{(2p+q).n}{2p.n} 4m^2\right)^{-\epsilon} ds,
 \label{BornPSpq}
\end{equation}
where $s = (2p+q)^2$.
The product of $N_{\rm Born}$, $ d\phi_{\rm Born}$,
and the function $ \mathcal{A}_{\rm Born}(p,q)$ in Eq.~\eqref{eq:ABorn}
defines a LO differential fragmentation function with general light-like vector $q$:
\begin{eqnarray}
\label{eq:NphiABorn}
N d\phi  \mathcal{A}_{\rm Born}(p,q)
&=& \frac{2(1-2 \epsilon) (4 \pi)^\epsilon\alpha_s^2}{\Gamma(2-\epsilon) N_c m}
[z (1-z)]^{-\epsilon}  \frac{(s-4m^2/z)^{-\epsilon}}{s^2}
\nonumber \\ 
&& \times 
\left[ 1 - \epsilon - 2 z(1-z) \frac{s (s-4m^2/z)} {(s-4m^2)^2} \right] ds,
\end{eqnarray}
where $s=(2p+q)^2$ and $z$ is the longitudinal momentum fraction
\begin{eqnarray}
z = \frac{(2p).n}{(2p+q).n} .
\label{eq:zpq}
\end{eqnarray}
If this measure is multiplied by a function of $s$ and integrated over $s$
from $4m^2/z$ to $\infty$, it defines a function of $z$.

In the calculation of the real NLO corrections to the fragmentation function,
it is convenient to have expressions for the Born squared amplitude
with a pair of uncontracted Lorentz indices.
They will be used to construct subtraction terms 
that cancel the ultraviolet and infrared divergences in the NLO corrections
point-by-point in the phase space. 
Such amplitudes with uncontracted indices cannot be expressed as a linear
combination of the contracted tensors in Eqs.~\eqref{eq:ggnnTraces}, but our 
prescription to extend $\gamma_5$ in $D$ dimensions can still be used.  
As will become clear later, contributions from subleading terms in $\epsilon$
always originate from the Laurent expansion of subtraction terms 
involving the Born squared amplitude with no Lorentz indices. Hence the study of 
alternative prescriptions for $\gamma_5$ can indeed ultimately be
reduced to its effects on the two expressions in Eqs~\eqref{eq:ggnnTraces}.
There are two useful choices 
for the uncontracted indices $\mu$ and $\nu$.  One choice is the  Lorentz indices
associated with the ends of the eikonal line.  The other choice is the 
Lorentz indices associated with the polarization vectors of 
the cut gluon line.  We will refer to those expressions as the {\it Born tensors}.

The Born tensor with Lorentz indices
associated with the eikonal line is
\begin{equation}
\mathcal{A}_{\rm eikonal}^{\mu \nu}(p,q) = 
\frac{(1-2\epsilon)(N_c^2-1) g_s^4  [(2p+q).n]^2}{2N_c m [(2p+q)^2]^2 (p.q)^2}
\left[ (2p.q)^2  T^{\mu \nu} - (2p+q)^2 l^\mu l^\nu \right] , 
\label{polarizedLOampeikonal}
\end{equation}
where $l^\mu$ and $T^{\mu \nu}$ are
\begin{subequations}
\begin{eqnarray}
l^\mu &= & 2p^\mu - \frac{2p.n}{(2p+q).n} (2p+q)^\mu ,
\\
T^{\mu \nu} & = & - g^{\mu \nu} + \frac{n^\mu (2p+q)^\nu + (2p+q)^\mu n^\nu}{(2p+q).n}.
\end{eqnarray}
\end{subequations}
They satisfy $l.n=0$ and $T^{\mu \nu}n_\nu=0$.\
Upon contracting $\mathcal{A}_{\rm eikonal}^{\mu \nu}$ with $-g_{\mu \nu}$,
we recover the Born squared amplitude in Eq.~\eqref{eq:ABorn}:
\begin{equation}
\label{eq:Borneikonal}
\mathcal{A}_{\rm Born}(p,q) 
= -g_{\mu \nu} \mathcal{A}_{\rm eikonal}^{\mu \nu}(p,q).
\end{equation}

The Born tensor with  Lorentz indices 
associated with the final-state gluon is
\begin{equation}
\mathcal{A}_{\rm gluon}^{\mu \nu}(p,q) =  
\frac{2 (N_c^2-1) g_s^4 [(2p+q).n]^2}{N_c m [(2p+q)^2]^2 (p.q)^2}
\sum_{i=1}^{4} C_i(z,p.q) T_{i}^{\mu \nu}(p,q),
\label{polarizedLOampgluon}
\end{equation}
where the tensors are
\begin{subequations}
\begin{eqnarray}
T_{1}^{\mu \nu} (p,q)& = & -g^{\mu \nu} +\frac{q^\mu n^\nu +  n^\mu q^\nu}{q.n}, \\
T_{2}^{\mu \nu} (p,q)& = & -g^{\mu \nu} +\frac{q^\mu p^\nu + p^\mu q^\nu }{p.q}, \\
T_{3}^{\mu \nu} (p,q)& = &   \left(p^\mu - \frac{p.q}{q.n} n^\mu \right) \left(p^\nu 
- \frac{p.q}{q.n} n^\nu \right), \\ 
T_{4}^{\mu \nu} (p,q)& = & q^\mu q^\nu.
\end{eqnarray}
\label{eq:Timunu}
\end{subequations}
Their coefficients are
\begin{subequations}
\begin{eqnarray}
C_1(z,p.q) & = & -2(1-z)(m^2 + p.q)\left[z p.q -2  (1-z)m^2 \right], \\
C_2(z,p.q) & = & \left[1-2\epsilon -2z(1-z) \right](p.q)^2-2 z (1-z)m^2p.q,  \\
C_3(z,p.q) & = &  4(1-z)^2(m^2 + p.q),\\
C_4(z,p.q) & = & z^2 p.q + (-1+2\epsilon +z^2)m^2 .
\end{eqnarray}
\end{subequations}
The argument $z$ of $C_i(z,p.q)$ is the 
longitudinal momentum fraction in Eq.~\eqref{eq:zpq},
which depends on $p$ and the arbitrary light-like vector $q$.
The tensors in Eqs.~\eqref{eq:Timunu} satisfy $T_i^{\mu \nu}q_\nu=0$
because of gauge invariance.
Upon contracting $\mathcal{A}_{\rm gluon}^{\mu \nu}$ with $-g_{\mu \nu}$,
we recover the Born squared amplitude in Eq.~\eqref{eq:ABorn}:
\begin{equation}
\label{eq:Borngluon}
\mathcal{A}_{\rm Born}(p,q) 
=  -g_{\mu \nu} \mathcal{A}_{\rm gluon}^{\mu \nu}(p,q).
\end{equation}

\section{Real NLO corrections}

\label{sec:NLOreal}

The real NLO corrections to the perturbative fragmentation function for 
$g \to Q \bar Q$, with the $Q \bar Q$ pair in a color-singlet $^1S_0$ state,
come from cut diagrams 
with two real partons in the final state.  The two partons can be 
two gluons or a light quark-antiquark pair ($q \bar q$).
Cut diagrams with two real gluons can be obtained 
from the four LO cut diagrams with a single real gluon,
such as the diagram in Figure~\ref{fig:cutdiagram},
by adding a gluon line that crosses the cut
and runs from any of the 6 colored lines on the left side of the cut 
to any of the 6 colored lines on the right side of the cut.
The additional gluon line can also be attached to the operator vertex, 
with the fragmenting gluon attached to the eikonal line.
The cut diagrams with a light $q \bar q$ pair can be obtained
from the four LO cut diagrams by replacing the real gluon line
that crosses the cut
by a virtual gluon that produces a $q \bar q$ pair that crosses the cut.

\subsection{Subtraction procedure}

\begin{figure}
\center
\includegraphics[scale=0.6]{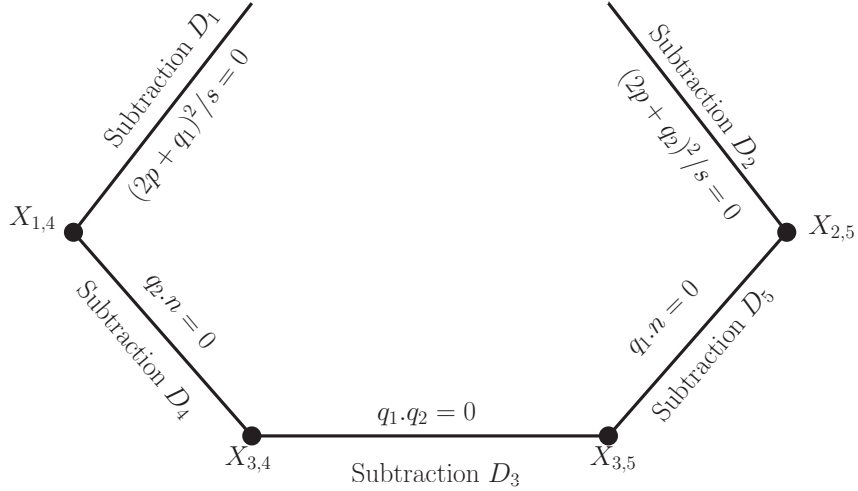}
\caption{Representation of the singular  regions for the integration 
of the real emission amplitude. Each line represents a specific limit, which is specified 
in terms of Lorentz invariants in the inner part of the figure. The relevant 
subtraction terms to extract the poles in each limit are also indicated. The phase-space regions
$X_{1,4}$ and $X_{2,5}$ yield a double pole $1/(\epsilon_{\textrm{IR}} \epsilon_{\textrm{UV}})$.
 The phase-space regions  $X_{3,4}$ and $X_{3,5}$ yield a double pole
 $1/\epsilon_{\textrm{IR}}^2$. }
\label{fig:poles}
\end{figure}

Each of the cut diagrams involves an integral over the phase space 
of the two real partons in the final state.  
The integrals diverge in several phase-space regions, yielding 
poles of both infrared (IR) and ultraviolet (UV) nature. 
Five (overlapping) boundaries in the phase-space can be associated 
with the singular behaviour of the integrand.
The boundaries can be defined in terms of Lorentz invariants.
We denote the momenta of both the $Q$ and $\bar Q$ by  $p$ 
and the momenta of the final-state partons 
(which can be gluons or a light quark and antiquark)
by $q_1$ and $q_2$.
The invariant mass of the four particles in the final state is 
$s=(2p+q_1+q_2)^2$.
The boundaries of the singular regions are represented in Figure~\ref{fig:poles}:
\begin{enumerate}  
\item the integration up to the boundary $(2p+q_1)^2/s = 0$ yields a UV pole,  
\item the integration up to the boundary $(2p+q_2)^2/s = 0$ yields a UV pole,  
\item the integration  up to the boundary $(q_1+q_2)^2/m^2 = 0$ yields an IR pole,  
\item the integration up to  the boundary $q_2.n/(q_1.n+q_2.n) = 0$ yields an IR pole,  
\item the integration up to the boundary $q_1.n/(q_1.n+q_2.n) = 0$ yields an IR pole.  
\end{enumerate}
The phase-space regions $X_{i,j}$ connecting two of the above boundaries
are associated with double poles, which can either be of pure infrared nature
(in the case of $X_{3,4}$ and $X_{3,5}$) or of mixed nature (in the case of $X_{1,4}$ and $X_{2,5}$). 

The real NLO contribution to the fragmentation  function can be expressed as
\begin{equation}
D_{g \to Q \bar Q}^{\rm (NLO,real)}(z) = 
N_{\rm CS} \int d\phi_{\rm real}(p,q_1,q_2) \mathcal{A}_{\rm real}\left(p,q_1,q_2\right),
\label{eq:real_emission}
\end{equation} 
where $N_{\rm CS}$ is the Collins-Soper prefactor in Eq.~\eqref{eq:overalfac},
$d\phi_{\rm real}$ is the product of the differential phase space 
for final-state partons with momenta $q_1$ and $q_2$
and the factor $2 \pi \delta(K.n - (2p+q_1+q_2).n)$
from the cut through the eikonal line,
and $\mathcal{A}_{\rm real}$ is the squared amplitude. 

Our strategy to extract the poles in the expression in Eq.~(\ref{eq:real_emission}) 
is to design a subtraction term $D_i$ for each of the five singular regions
in Figure~\ref{fig:poles} whose integral over that region has poles 
that match those of the integral of $ \mathcal{A}_{\rm real}$.
The contribution to the fragmentation  function in Eq.~\eqref{eq:real_emission} 
can be expressed as
\begin{eqnarray}
D_{g \to Q \bar Q}^{\rm (NLO,real)}(z) &=& 
N_{\rm CS} \int d\phi_{\rm real}(p,q_1,q_2) 
\left[ \mathcal{A}_{\rm real}\left(p,q_1,q_2\right) 
- \sum_{i=1}^5 D_i(p,q_1,q_2)
\right]  
\nonumber \\
&&+ \sum_{i=1}^{5} N_{\rm CS} \int d\phi_{\rm real}(p,q_1,q_2)   D_i(p,q_1,q_2).
\label{eq:real_emission_subtracted}
\end{eqnarray}
The subtraction terms $D_i$ are designed so that
the integral in the first term is finite and can be evaluated in $D=4$ dimensions. 
The integrals in the second term are evaluated in $D=4-2\epsilon$ dimensions,
so the UV and IR divergences appear as poles in $\epsilon$.  
The construction of the subtraction terms $D_i$
is described in Sections~\ref{sec:realsubUV} and \ref{sec:realsubIR},
where we follow closely the subtraction procedure introduced by 
Catani and Seymour~\cite{Catani:1996vz}.
The analytic integration of the subtraction terms to obtain the poles in $\epsilon$
is described in Sections~\ref{sec:realintUV} and \ref{sec:realintIR},
where we again follow closely the procedure introduced in Ref.~\cite{Catani:1996vz}.

\subsection{Subtractions for UV and mixed poles}
\label{sec:realsubUV}

The UV poles in the real NLO contribution to the fragmentation  function
are matched by the integrals of the subtraction terms 
$D_1$ associated with the limit $(2p+q_1)^2/s \rightarrow 0$
and $D_2$ associated with the limit $(2p+q_2)^2/s \rightarrow 0$.
The total invariant mass $s$, the invariant mass $s_i$ for the system 
consisting of the $Q \bar Q$ pair and the parton of momentum $q_i$,
and the longitudinal momentum fraction $y_i$ for that system are
\begin{equation}
\label{eq:yi-pq1q2}
s = (2p+q_1+q_2)^2,   \qquad
s_i = (2p+q_i)^2,   \qquad y_i = \frac{(2p+q_i).n}{(2p+q_1+q_2).n}.
\end{equation}
Our subtraction term $D_i$ associated with the limit $s_i/s \rightarrow 0$
includes a factor of $\mathcal{A}_{\rm eikonal}^{\mu \nu}(p,q_i)$,
where $\mathcal{A}_{\rm eikonal}^{\mu \nu}$ is the Born tensor 
defined in Eq.~(\ref{polarizedLOampeikonal})
whose Lorentz indices $\mu$ and $\nu$ are associated with the eikonal line.
The factor $\mathcal{A}_{\rm eikonal}^{\mu \nu}$ can be interpreted 
as arising from the fragmentation of a gluon with longitudinal momentum 
$y_iK.n$ into a $Q \bar Q$ pair with longitudinal momentum $z K.n$ 
via the radiation of a gluon of momentum $q_i$.

The subtraction terms $D_1$ and $D_2$ are given by
\begin{equation}
D_i(p,q_1,q_2) = \frac{4 \pi \alpha_s \mu^{2 \epsilon}}{s} V^\txtsm{UV}_{\mu \nu}(y_i,l_i) 
  \frac{1}{y_i^2} \mathcal{A}_{\rm eikonal}^{\mu \nu}(p,q_i ) , \quad i=1,2,
\label{eq:sub1and2}
\end{equation}
where the kernel $V^\txtsm{UV}_{\mu \nu}(y_i,l_i)$ is defined by
\begin{equation}
V^{\txtsm{UV}}_{\mu \nu}(y,l) = 2N_c \Bigg[ \left(  \frac{y}{1-y} + 
y \left(1 - y \right) \right) (-g_{\mu \nu})
- 2 (1- \epsilon) 
 \frac{1-y}{y}
 \frac{l_\mu l_\nu }{l^2} \Bigg].
\end{equation}
The 4-vectors $l_1$ and $l_2$ appearing as the second argument of
$V^\txtsm{UV}_{\mu \nu}(y_i,l_i)$ in Eq.~\eqref{eq:sub1and2} 
are defined by
\begin{subequations}
\begin{eqnarray}
l_1^\mu &=& q_2^\mu - \frac{q_2.n}{(2p+q_1).n}  (2p+q_1)^\mu,
\\
l_2^\mu &=& q_1^\mu -  \frac{q_1.n}{(2p+q_2).n} (2p+q_2)^\mu.
\end{eqnarray}
\label{eq:l1l2}
\end{subequations}
These 4-vectors are orthogonal to $n$:  $l_i . n = 0$.

\subsection{Integrals with UV and mixed poles}
\label{sec:realintUV}
 
 Explicit expressions for the poles in the integral of 
 the subtraction term $D_i$ ($i=1,2$)
 can be obtained by carrying out the integration over the
$(3-2\epsilon)$-dimensional slice 
 associated with a fixed value of $s_i = (2p+q_i)^2$. A convenient  
 decomposition of the phase-space measure is
 derived in Appendix~\ref{sec:PhaseSpace}:
 \begin{equation}
\label{eq:Ndphi_UV}
N_{\rm CS} d\phi_{\rm real}(p,q_1,q_2) = 
N_{\rm Born}(p, q_i) d\phi_{\rm Born}(p,q_i) \;  d \phi^{(i)} (p,q_1,q_2).
\end{equation}
The prefactor $N_{\rm Born}(p,q_i)$ is
defined in Eq.~(\ref{eq:overalfacBORN}).
The factor $d\phi_{\rm Born}(p,q_i)$, which is differential in $s_i$,
is defined in Eq.~(\ref{BornPSpq}).
The measure $d \phi^{(i)} $ for integration over the slice with fixed $s_i$ is 
\begin{equation}
d \phi^{(i)}(p,q_1,q_2) =  
\frac{1}{4(2\pi)^{3-2\epsilon}} (s - s_i/y_i)^{-\epsilon}  ds \; 
y_i^{1-\epsilon} (1-y_i)^{-\epsilon} dy_i\;  d\Omega_\perp ,
\label{eq:def_dphi1_UV}
\end{equation}
where $d\Omega_\perp$ is the transverse angular measure
whose integral is $2 \pi^{1-\epsilon}/\Gamma(1-\epsilon)$.
The differential variables $s$, $s_i$, and $y_i$ are defined as functions of 
$p$, $q_1$, and $q_2$ in Eqs.~\eqref{eq:yi-pq1q2}.
The range of $y_i$ is from $z$ to $1$.
The range of $s_i$ is from $4m^2/(z/y_i)$ to $\infty$,
and the range of $s$ is from $s_i/y_i$ to $\infty$.

To carry out the integration over the transverse angles in $\Omega_\perp$, 
we observe that the 4-vectors $l_1^\mu$ and  $l_2^\mu$ defined in Eq.~\eqref{eq:l1l2}
are orthogonal to $n^\mu$, so Lorentz invariance implies
\begin{equation}
\int d\Omega_\perp \frac{l^\mu_i l^\nu_i}{l_i^2} = 
A \left( - g^{\mu \nu} 
+ \frac{n^\mu (2p+q_i)^\nu + n^\nu (2p+q_i)^\mu }{(2p+q_i).n} \right) 
+ B n^\mu n^\nu ,
\label{azimuthal_int_D12}
\end{equation}
where $A$ and $B$ are functions of $s_i$, $y_i$ and $u$.
Because of gauge invariance, the amplitude $\mathcal{A}_{\rm eikonal}^{\mu \nu}$ 
is orthogonal to $n^\mu$ and $n^\nu$, 
so that only the term $A (-g^{\mu\nu})$ survives after contracting the tensor 
on the right side of Eq.~(\ref{azimuthal_int_D12})
with  $\mathcal{A}_{\rm eikonal}^{\mu \nu}$. 
We can determine the coefficient $A$ by contracting 
both sides of 
Eq.~\eqref{azimuthal_int_D12} by $g_{\mu\nu}$:
\begin{equation}
\label{eq:A}
A = - \frac{\pi^{1-\epsilon}}{\Gamma(2- \epsilon)}.
\end{equation}
After integrating over the angles in 
$\Omega_\perp$, one can make the replacement
\begin{equation}
\int d\Omega_\perp  V^{\txtsm{UV}}_{\mu \nu}(y_i,l_i)  
\longrightarrow  \frac{2\pi^{1-\epsilon}}{\Gamma(1-\epsilon)}
\hat P_{gg}^{\rm (real)}(y_i) ( -g_{\mu \nu}),
\label{int_sphere}
\end{equation}
where $\hat P_{gg}^{\rm (real)}(y)$ is the real-gluon
contribution to the Altarelli-Parisi splitting function for $g \rightarrow g$ 
without any regularization of the pole at $y=1$:
\begin{equation}
\label{eq:Pgg-realdiv}
\hat P_{gg}^{\rm (real)}(y) = 2N_c
\left[ \frac{y}{1-y} + \frac{1-y}{y} +y(1-y)  \right] .
\end{equation}

The contraction of $-g_{\mu \nu}$ in Eq.~(\ref{int_sphere}) with 
the Born tensor $ \mathcal{A}_{\rm eikonal}^{\mu \nu}(p,q_i)$ 
gives the Born squared amplitude $\mathcal{A}_\txtsm{Born}(p,q_i)$ 
obtained from Eq.~\eqref{eq:ABorn} by replacing $q$ by $q_i$.
The UV pole can be made explicit by integrating analytically over the variable $s$:
\begin{eqnarray}
\label{lambda-2-integrated}
 N_{\rm CS} \int d\phi_{\rm real}  D_i (p,q_1,q_2)  &=& 
 \frac{\Gamma(1+\epsilon)}{\epsilon_{\textrm{UV}}}
 \frac{\alpha_s}{4\pi} \left(\frac{\pi \mu^2}{m^2} \right)^\epsilon
\int_z^1 \frac{dy_i}{y_i} (1-y_i)^{-\epsilon}
\hat P_{gg}^{\rm (real)}(y_i)  
\nonumber \\ 
& & \times
\int N d\phi \mathcal{A}_{\rm Born}( p, q_i) \;
(s_i/4 m^2)^{-\epsilon} ,
\end{eqnarray}
where $N d\phi \mathcal{A}_{\rm Born}( p, q_i)$ is the 
LO differential fragmentation function
obtained from Eq.~\eqref{eq:NphiABorn} by replacing $q$ by $q_i$.
The variables $s$ and $z$ in Eq.~\eqref{eq:NphiABorn}
are replaced by $s_i$ and $z/y_i$.
The IR pole associated with the $y_i=1$ endpoint
can be extracted by applying the plus prescription:
\begin{eqnarray}
\label{AP-reg}
(1-y)^{-\epsilon} \hat P_{gg}^{\rm (real)}(y) &=& P_{gg}^{\rm (real)}(y) -2N_c \delta(1-y)\frac{1}{\epsilon_\textrm{IR}} 
\\ \nonumber & &
 -  2N_c  \Bigg[ \left( \frac{\log(1-y)}{1-y} \right)_+ 
 + \left( \frac{1}{y}+y(1-y)-2 \right) \log(1-y) \Bigg] \epsilon
+ \mathcal{O}(\epsilon^2),
\end{eqnarray}
where $ P_{gg}^{\rm (real)}(y)$  is the real-gluon
contribution to the Altarelli-Parisi splitting function for $g \rightarrow g$:
\begin{equation}
 P_{gg}^{\rm (real)}(y)= 2N_c
 \left[ \frac{y}{(1-y)_+} + \frac{1-y}{y}  + y(1-y)  \right].
\end{equation}

With the use of Eq.~(\ref{AP-reg}),  the expression in 
Eq.~(\ref{lambda-2-integrated}) can be expressed as 
the sum of a term with a double pole, a term with a single pole, and a finite remainder:
\begin{equation}
\sum_{i=1}^{2} N_{\rm CS} \int d\phi_{\rm real} D_i(p,q_1,q_2) =
\frac{\alpha_s}{2\pi} \Gamma(1+\epsilon) \left(\frac{\pi \mu^2}{m^2} \right)^\epsilon
[ I_2(z)+ I_1(z) + I_0(z)],
\label{D1D2_integrated}
\end{equation}
where the functions $I_n(z)$ are 
\begin{subequations}
\begin{eqnarray}
 I_2(z) &=&
- \frac{2N_c}{\epsilon_{\textrm{UV}}\epsilon_{\textrm{IR}}} 
\left[ D^{\rm (LO)}_{g \to Q \bar Q}(z) - \epsilon D_{\log}(z) \right]  ,
 \label{eq:Idouble} 
 \\
 I_1(z)  &=&  \frac{1}{\epsilon_{\textrm{UV}}}  \int_z^1 \frac{dy}{y}   
P_{gg}^{\rm (real)}(y) D^{\rm (LO)}_{g \to Q \bar Q}(z/y), 
\label{eq:Isingle}  
\\
\label{eq:remainder}
I_0(z) & =&   -  N_c D_{\log^2}(z/y) 
 - \int_z^1 \frac{dy}{y}  P_{gg}^{\rm (real)}(y)  D_{\log}(z/y) 
\\ & &   -2 N_c 
\int_z^1 \frac{dy}{y}    
\bigg[  \left( \frac{\log(1-y)}{1-y} \right)_+
+ \left(\frac{1}{y}+y(1-y)-2 \right)\log(1-y) \bigg] 
D^{\rm (LO)}_{g \to Q \bar Q}(z/y) . \nonumber 
\end{eqnarray}
\end{subequations}
The LO fragmentation function $D^{\rm (LO)}_{g \to Q \bar Q}(z)$ is defined in
Eq.~\eqref{eq:DLOeps}.  The functions $D_{\log}(z)$ and $D_{\log^2}(z)$ 
are defined by
\begin{subequations}
\label{eq:Dlog12}
\begin{eqnarray}
D_{\log}(z) &=& \int N d\phi  \mathcal{A}_{\rm Born}(p, q)  
\log(s/4m^2) ,
\\
D_{\log^2}(z) &=& \int N d\phi  \mathcal{A}_{\rm Born}(p, q)  
\log^2(s/4m^2) ,
\end{eqnarray}
\end{subequations}
where the measure $N d\phi  \mathcal{A}_{\rm Born}(p,q)$,
which is differential in $s=(2p+q)^2$, is given in Eq.~\eqref{eq:NphiABorn}.
These functions appear in Eqs.~\eqref{eq:Isingle} and \eqref{eq:remainder}
with argument $z/y$.
In Eqs.~\eqref{eq:Idouble} and \eqref{eq:Isingle}, 
there are terms of order $\epsilon^0$ from
the expansion of $D^{\rm (LO)}_{g \to Q \bar Q}(z)$  to order $\epsilon^2$
and the expansion of  $D_{\log}(z)$ to order $\epsilon$.
These expansions are not actually needed, because the canceling 
poles in $\epsilon$ will also be expressed in terms of the functions
$D^{\rm (LO)}_{g \to Q \bar Q}(z)$ and $D_{\log}(z)$.

\subsection{Subtractions for IR poles}
\label{sec:realsubIR}

The IR poles in the real NLO contribution to the fragmentation  function
are matched by the subtraction terms $D_3$, $D_4$, and $D_5$ 
associated with the limits $q_1.q_2 \to 0$, $q_2.n \to 0$, 
and  $q_1.n \to 0$, respectively. 
The expressions for these subtraction terms can be made more compact by introducing a
light-like 4-vector $\tilde{q}$ that has the same longitudinal momentum as $q_1+q_2$:
\begin{equation}
\tilde{q}^\mu= (q_1+q_2)^\mu -  \frac{q_1.q_2}{(q_1+q_2).n} n^\mu.
\end{equation}
It satisfies $\tilde q^{\,2} = 0$ and $\tilde q.n = (q_1+q_2).n$.
It is also convenient to introduce variables $\tilde{s}$, $u$, and $\lambda$
defined by
\begin{eqnarray}
\tilde{s} = (2p+\tilde{q})^2, \qquad
u = \frac{q_2.n}{(q_1 + q_2).n}, \qquad
\lambda =\frac{(q_1+q_2)^2}{4m^2}.
\label{eq:sulambda-q1q2}
\end{eqnarray}
Our subtraction terms $D_3$, $D_4$, and $D_5$
include a factor of $\mathcal{A}_{\rm gluon}^{\mu \nu}(p,\tilde q)$,
where $\mathcal{A}_{\rm gluon}^{\mu \nu}$
is the Born tensor defined in Eq.~(\ref{polarizedLOampgluon})
whose Lorentz indices $\mu$ and $\nu$ are associated with the final-state gluon.
The factor $\mathcal{A}_{\rm gluon}^{\mu \nu}$ can be interpreted as arising 
from the fragmentation of a gluon with longitudinal momentum $K.n$
into a $Q \bar Q$ pair with longitudinal momentum $zK.n$ 
via the radiation of a gluon of momentum $\tilde q$.

The subtraction terms $D_4$ and $D_5$ are defined by
\begin{equation}
\label{eq:D4D5}
D_i(p,q_1,q_2) = \frac{2 \pi \alpha_s \mu^{2 \epsilon}}{m^2} 
V^{\txtsm{IR}(i)}(p,q_1,q_2) 
( -g_{\mu \nu} ) \mathcal{A}_ {\rm gluon}^{\mu \nu}(p,\tilde{q}) , \quad i=4,5,
\end{equation}
where the kernels $V^{\txtsm{IR}(i)}(p,q_1, q_2)$ are 
\begin{subequations}
\begin{eqnarray}
V^{\txtsm{IR}(4)}(p,q_1,q_2) &=&   
\frac{N_c \tilde{s}}{u(u + \lambda)[\tilde{s}+4m^2\lambda/(1-z)]} , 
 \\ 
 V^{\txtsm{IR}(5)}(p,q_1,q_2) &=&
 \frac{N_c \tilde{s}}{(1-u)(1-u + \lambda)[\tilde{s}+4m^2\lambda/(1-z)]} .
\end{eqnarray}
\end{subequations}

The subtraction term $D_3$ matches the IR poles originating from collinear partons in the final state. It can be expressed in the form
\begin{equation}
\label{eq:D3}
D_3(p,q_1,q_2) = \frac{4 \pi \alpha_s \mu^{2 \epsilon}}{(q_1+q_2)^2} 
\left[ V^{gg}_{\mu \nu}(q_1, q_2) + V^{q\bar q}_{\mu \nu}(q_1, q_2) \right]
\mathcal{A}_ {\rm gluon}^{\mu \nu}(p,\tilde{q} ) .
\end{equation}
The kernel $V^{gg}_{\mu \nu} + V^{q\bar q}_{\mu \nu}$ has been split into two terms
associated with collinear gluons and collinear quarks.
It is convenient to introduce a 4-vector ${\breve q}(u)$ whose components are
\begin{equation}
\label{eq:qu}
{\breve q}(u)^\mu = u {q_2}^\mu -(1-u) {q_1}^\mu.
\end{equation}
The kernels
associated with collinear gluons and collinear quarks are
\begin{subequations}
\label{eq:VggVqq}
\begin{eqnarray}
V^{gg}_{\mu \nu}(q_1,q_2) & =&  2N_c \Bigg[ \left( 
 \frac{1-u}{u+\lambda} + \frac{u}{1-u+\lambda} 
 \right) (-g_{\mu \nu})
+ (1- \epsilon) 
\frac{1}{1+\lambda }  \frac{{\breve q}(u)_\mu {\breve q}(u)_\nu}{q_1.q_2} \Bigg],
\\
V^{q \bar q}_{\mu \nu}(q_1,q_2)  &=&2 T_F n_f
\left[ \frac{1}{1+\lambda}  (- g_{\mu \nu})  
- \frac{2}{1+\lambda }  \frac{{\breve q}(u)_\mu {\breve q}(u)_\nu}{q_1.q_2}  \right] ,
\end{eqnarray}
\end{subequations}
where $T_F= \frac12$ is the trace of the square of a generator 
for the fundamental representation.

\subsection{Integrals with IR poles}
\label{sec:realintIR}

 Explicit expressions for the poles in the subtraction terms $D_3$,
 $D_4$, and $D_5$
 can be obtained by carrying out the phase-space integration over the 
$(3-2\epsilon)$-dimensional slice 
 associated with a fixed value of $\tilde{s}=(2p+\tilde{q})^2$. 
A convenient decomposition of the phase-space measure is 
 derived in Appendix~\ref{sec:PhaseSpace}:
\begin{equation}
\label{eq:Ndphi_IR}
N_{\rm CS} d \phi_{\rm real}(p,q_1,q_2) = 
N_{\rm Born}(p,\tilde q)d\phi_{\rm Born}(p,\tilde q) \;d \tilde \phi(p,q_1,q_2).
\end{equation} 
The prefactor $N_{\rm Born}(p,\tilde q)$, which is
defined in Eq.~(\ref{eq:overalfacBORN}), coincides with $N_{\rm CS}$.
The factor $d\phi_{\rm Born}(p,\tilde q)$, 
which is differential in $\tilde s$,  is defined in Eq.~\eqref{BornPSpq}.
The measure $d \tilde \phi$ for integration over the slice with fixed $\tilde s$ is
\begin{equation}
d \tilde \phi(p,q_1,q_2) =\frac{(4m^2)^{1-\epsilon}}{4 (2\pi)^{3-2\epsilon}}  
u^{-\epsilon} (1-u)^{-\epsilon} du\; \lambda^{-\epsilon} d\lambda\; d\Omega_\perp ,
\label{eq:dtildephi_IR}
\end{equation}
where $d\Omega_\perp$ is the transverse angular measure. 
The differential  variables $\tilde s$,
$u$, and $\lambda$ are defined as functions of $p$, $q_1$, and $q_2$
in  Eqs.~\eqref{eq:sulambda-q1q2}.
The range of $\tilde s$ is from $4m^2/z$ to $\infty$.
The range of $\lambda$ is from $0$ to $\infty$,
and the range of $u$ is from $0$ to $1$.

In the expressions for $D_4$ and $D_5$ in
Eq.~\eqref{eq:D4D5}, the contraction of $-g_{\mu \nu}$
with $ \mathcal{A}_ {\rm gluon}^{\mu \nu}(p,\tilde{q})$
gives the Born squared amplitude $\mathcal{A}_\txtsm{Born}(p,\tilde{q} )$
obtained from Eq.~\eqref{eq:ABorn} by replacing $q$ by $\tilde q$.
In the expression for $D_3$ in Eq.~\eqref{eq:D3}, 
the Born tensor $ \mathcal{A}_ {\rm gluon}^{\mu \nu}(p,\tilde{q})$
is contracted with the tensors $V^{gg}_{\mu \nu}$ 
and $V^{q \bar q}_{\mu \nu}$ defined in Eq.~\eqref{eq:VggVqq}.
A factor of $ \mathcal{A}_ {\rm Born}(p,\tilde{q})$
appears only after  integrating over the 
transverse angles in $\Omega_\perp$.
To carry out that integration,
we observe that the 4-vector $\breve{q}(u)$ defined in Eq.~\eqref{eq:qu}
is orthogonal to $\tilde{q}$, so Lorentz invariance implies
\begin{equation}
\int d\Omega_\perp \frac{\breve{q}(u)^\mu \breve{q}(u)^\nu}{q_1.q_2} = 
C \left( - g^{\mu \nu} + \frac{n^\mu \tilde{q}^\nu + n^\nu \tilde{q}^\mu }{n.\tilde{q}} 
\right) + D \tilde{q}^\mu \tilde{q}^\nu,
\label{azimuthal_int_D345}
\end{equation}
where the coefficients $C$ and $D$ are functions of $q_1.q_2$ and $u$.
Because of gauge invariance, the Born tensor 
$\mathcal{A}_ {\rm gluon}^{\mu \nu}(p,\tilde{q} )$ 
is orthogonal to $\tilde{q}^\mu$ and $\tilde{q}^\nu$, 
so only the term $C (-g_{\mu\nu})$ survives after contracting the tensor
on the left side of Eq.~(\ref{azimuthal_int_D345}) 
with $\mathcal{A}_ {\rm gluon}^{\mu \nu}(p,\tilde{q} )$.
We can determine the coefficient $C$ by contracting 
both sides of Eq.~\eqref{azimuthal_int_D345} by $g_{\mu\nu}$:
\begin{equation}
\label{eq:C}
C = \frac{2 \pi^{1-\epsilon}}{\Gamma(2- \epsilon)} u (1-u) .
\end{equation}

The IR poles can be made explicit by  integrating 
over the variables $u$ and $\lambda$.
The integral over $\lambda$ can be evaluated analytically, 
and it gives a pole in $\epsilon$.  
The integral over $u$ gives a second pole in $\epsilon$.
After isolating the term that gives the pole,
the integrand can be expanded in powers of $\epsilon$
 and then integrated over $u$.
The resulting expressions for the integrals of $D_4$ and $D_5$ are the same:
\begin{equation}
N_{\rm CS}\int d\phi_{\rm real} D_i(p,q_1,q_2) = 
\frac{\alpha_s }{2\pi} \Gamma(1+\epsilon)\left(\frac{\pi \mu^2}{m^2} \right)^\epsilon
\int N d\phi \mathcal{A}_{\rm Born}(p, \tilde{q}) \;
 \mathcal{V} (p, \tilde{q}),  \quad i=4,5,
 \label{eq:integralD45}
\end{equation} 
where $N d\phi \mathcal{A}_{\rm Born}( p, \tilde q)$ is the 
LO differential fragmentation function
obtained from Eq.~\eqref{eq:NphiABorn} by replacing $q$ by $\tilde q$.
The function $ \mathcal{V} (p, \tilde{q}) $ includes all the poles in $\epsilon$: 
\begin{equation}
\label{def_V}
\mathcal{V}(p, \tilde{q}) = N_c \left[ \frac{1}{2}
\left( \frac{1}{\epsilon_{\rm{IR}}} - \log \frac{(1-z)\tilde{s}}{4m^2} \right)^2
-\textrm{Li}_2\bigg( 1-\frac{4m^2}{(1-z)\tilde{s}} \bigg)+\frac{\pi^2}{6}  \right] .
  \end{equation} 
After integrating over $\tilde s$, the integral of $D_4+D_5$  reduces to
\begin{eqnarray}
\label{eq:NdPhiD45}
&&\sum_{i=4,5} N_{\rm CS} \int d\phi_{\rm real} D_i(p,q_1,q_2) \nonumber \\
&&= 
\frac{\alpha_s N_c}{2\pi}
\Gamma(1+\epsilon)\left(\frac{\pi \mu^2}{m^2} \right)^\epsilon
\Bigg\{ 
 \left[ \frac{1}{\epsilon_{\rm{IR}}^2} -
\frac{2}{\epsilon_{\rm{IR}}}\log(1-z)  +\log^2(1-z) +\frac{\pi^2}{3}  \right]
 D^{\rm (LO)}_{g \to Q \bar Q}(z)
\nonumber \\
&& \hspace{4.5cm}
- 2 
\left(\frac{1}{\epsilon_{\rm{IR}}} - \log(1-z) \right) D_{\log}(z)
+ D_{\log^2}(z) - 2 D_{\rm Li}(z) \Bigg\}.
\end{eqnarray} 
The LO fragmentation function $D^{\rm (LO)}_{g \to Q \bar Q}(z)$ is  defined in
Eq.~\eqref{eq:DLOeps} and the functions $D_{\log}(z)$ and $D_{\log^2}(z)$ 
are defined in Eqs.~\eqref{eq:Dlog12}.
The function $D_{\rm Li}(z)$ is defined by
\begin{eqnarray}
\label{eq:DLi}
D_{\rm Li}(z) &=& \int N d\phi  \mathcal{A}_{\rm Born}(p, q)  
\textrm{Li}_2\bigg( 1-\frac{4m^2}{(1-z)s} \bigg) ,
\end{eqnarray}
where the measure $N d\phi  \mathcal{A}_{\rm Born}$,
which is differential in $s = (2p+q)^2$, is given in Eq.~\eqref{eq:NphiABorn}.
In Eq.~\eqref{eq:NdPhiD45},
there are terms of order $\epsilon^0$ from
the expansion of $D^{\rm (LO)}_{g \to Q \bar Q}(z)$  to order $\epsilon^2$
and the expansion of  $D_{\log}(z)$ to order $\epsilon$.
These expansions are not actually needed, because the canceling 
poles in $\epsilon$ will also be expressed in terms of the functions
$D^{\rm (LO)}_{g \to Q \bar Q}(z)$ and $D_{\log}(z)$.

In the expression for the integral of $D_3$, the IR poles
appear in a multiplicative constant factor that can be separated into 
contributions from gluons and quarks:
\begin{equation}
N_{\rm CS}\int d\phi_{\rm real} D_3(p, q_1,q_2) = 
\frac{\alpha_s }{2\pi} \Gamma(1+\epsilon)\left(\frac{\pi \mu^2}{m^2} \right)^\epsilon
\left(  \mathcal{V}^{gg}+\mathcal{V}^{q \bar q} \right)
\int N d\phi \mathcal{A}_{\rm Born}(p, \tilde{q}).
 \label{eq:integralD3}
\end{equation} 
The factors $ \mathcal{V}^{gg}$ and $\mathcal{V}^{q \bar q}$
are integrals over $\lambda$ and $u$ that can be evaluated analytically.
Their Laurent expansions to order $\epsilon^0$ are
\begin{subequations}
\label{def_VggVqq}
\begin{eqnarray}
\mathcal{V}^{gg}  & = & 
N_c \left[ 
\frac{1}{\epsilon_{\rm{IR}}^2} + \frac{11}{6 \epsilon_{\rm{IR}}} 
+ \frac{103}{18}-\frac{\pi^2}{3} \right] ,
\label{def_Vgg}
\\
\mathcal{V}^{q \bar q}  & = & T_F n_f 
\left[-\frac{2}{3 \epsilon_{\rm{IR}} } -\frac{10}{9}  \right].
\label{def_Vqq}
\end{eqnarray}
\end{subequations}
After integrating over $\tilde s$, the integral of $D_3$ reduces to
\begin{equation}
\label{eq:D3int}
N_{\rm CS}\int d\phi_{\rm real} D_3(p, q_1,q_2) = 
\frac{\alpha_s }{2\pi} \Gamma(1+\epsilon)\left(\frac{\pi \mu^2}{m^2} \right)^\epsilon
\left(  \mathcal{V}^{gg}+\mathcal{V}^{q \bar q} \right) D^{\rm (LO)}_{g \to Q \bar Q}(z).\end{equation} 
The poles in $\epsilon$ in $ \mathcal{V}^{gg}+\mathcal{V}^{q \bar q}$
will be cancelled by another term proportional to
$D^{\rm (LO)}_{g \to Q \bar Q}(z)$,  
leaving only the  terms of order $\epsilon^0$.

\section{Virtual NLO corrections}

\label{sec:NLOvirtual}

The virtual NLO corrections to the perturbative fragmentation function for 
$g \rightarrow Q \bar Q$,
with the $Q \bar Q$ pair in a color-singlet $^1S_0$ state, 
come from cut diagrams with one loop on either 
the right side or the left side of the cut. 
Loop diagrams on one side of the cut can be obtained 
from the LO diagrams by adding a gluon line connecting any pair 
of the 6 colored lines, by adding a loop correction to the propagator 
of the fragmenting gluon, or by adding a loop correction to the propagator 
of the virtual heavy quark.
There are additional loop diagrams in which the heavy quark line is attached 
to the eikonal line by both the fragmenting gluon that attaches to the 
end of the eikonal line and by a second gluon line, 
with the gluon that crosses the cut attached to either the fragmenting gluon 
or the eikonal line.

As in the LO cut diagrams,
we denote the momenta of both the $Q$ and $\bar Q$ by $p$
and the momentum of the final-state gluon by $q$.
We denote the loop momentum by $l$.
The sum of the virtual one-loop cut diagrams at order $\alpha_s^3$ defines 
a function $\mathcal{A}_{\rm virtual}\left(p,q,l\right)$. 
The virtual NLO contribution to the fragmentation function can be expressed as
\begin{equation}
D_{g \to Q \bar Q}^{\rm (virtual)}(z) = 
N_{\rm CS} \int d\phi_{\rm Born}
 \int \frac{d^Dl}{(2\pi)^D} \mathcal{A}_{\rm virtual}\left(p,q,l\right) \, ,
\label{eq:virtual_emission}
\end{equation} 
where $N_{\rm CS}$ is the Collins-Soper prefactor in Eq.~(\ref{eq:overalfac}) 
and $d\phi_{\rm Born}$
is the phase-space measure in Eq.~\eqref{eq:BornPS}. 

By means of standard tensor reduction techniques,
the integral over the loop momentum $l$ in Eq.~(\ref{eq:virtual_emission})
can be reduced to a sum of one-loop scalar integrals
whose numerators are simply 1. 
Our procedure to apply this 
reduction is implemented with the use of the Mathematica package 
FeynCalc~\cite{Mertig:1990an}.
The denominators of the scalar integrals come from Feynman propagators
with mass $m$, massless Feynman propagators, and eikonal propagators 
of the form $i/[(l+P).n + i \epsilon]$, 
where $P$ is a linear combination of $p$ and $q$.
A product of eikonal propagators can be reduced algebraically to 
a linear combination of single eikonal propagators.
A scalar integral with only Feynman propagators
is a function of the invariant mass $s =(2p+q)^2$.
A scalar integral with one eikonal propagator is a function of $s$ 
and the momentum fraction $z=(2p).n/(2p+q).n$.
We need the Laurent expansion in $\epsilon=(4-D)/2$ 
for each scalar integral to order $\epsilon^0$. 
Our results for the scalar integrals with only Feynman propagators
are in agreement with results available in the 
literature~\cite{Ellis:2007qk,vanOldenborgh:1990yc}.
The Laurent expansions for the scalar integrals 
with a single eikonal propagator can be evaluated analytically,
with the finite terms order of $\epsilon^0$ expressed in terms of dilogarithms.
For some of the integrals, the analytic expressions in terms of dilogarithms
are very complicated, so the finite terms might as well be 
expressed in terms of finite integrals that can be evaluated numerically.
Our results for the poles in $\epsilon$ 
in the scalar integrals with one eikonal propagator
are given in Appendix~\ref{sec:LoopIntegrals}.

In all the poles in $\epsilon$ from the loop integral in 
Eq.~(\ref{eq:virtual_emission}), the Born squared amplitude 
$\mathcal{A}_{\textrm{Born}}(p,q)$ appears as a multiplicative factor.  
The virtual NLO corrections to the fragmentation function
can therefore be expressed as
\begin{eqnarray}
\label{virtual_correction0}
 D_{g \to Q \bar Q}^{\rm (virtual)}(z)
 &=& \frac{\alpha_s}{2 \pi}  
 \Gamma(1+\epsilon) \left(\frac{\pi \mu^2}{m^2}\right)^\epsilon 
  N_\textrm{CS} \int d\phi_{\rm Born}   
  \left[f_\textrm{pole}(p,q) \mathcal{A}_{\textrm{Born}}(p,q)
+ \mathcal{A}_{\textrm{finite}}(p,q) \right],
\nonumber \\
\end{eqnarray}
where $f_\textrm{pole}(p,q)$ has only poles in $\epsilon$
and  $\mathcal{A}_{\textrm{finite}}(p,q)$ is a finite function of 
$s =  (2p+q)^2$ and $z$.
The terms in $f_\textrm{pole}(p,q)$ can be organized
to make their cancellation against the poles from other contributions 
of the NLO correction more transparent:  
\begin{equation}
\label{one-loop-div}
f_\textrm{pole}(p,q)  = 
\mathcal{U}^g + \mathcal{U}^Q(s) + 2\mathcal{U}^{gQ\bar Q}+ \mathcal{U}^{\rm eikonal}
+ \mathcal{M}(s) + \mathcal{S}_1 + \mathcal{S}_2(s,z) . 
\end{equation}
There are four terms in Eq.~(\ref{one-loop-div}) with only UV poles:
\begin{subequations}
\begin{eqnarray}
\mathcal{U}^g & = & \left( \frac{5}{3} N_c
- \frac{4}{3}T_F n_f \right)\frac{1}{\epsilon_\textrm{UV}} ,
\\
\mathcal{U}^Q(s) & = &  C_F\frac{1}{\epsilon_\textrm{UV} }
 \left(  \frac{12m^2}{s-4m^2}  -1 \right),
 \\
\mathcal{U}^{gQ \bar Q} & = &  \left( N_c + C_F  \right)\frac{1}{\epsilon_\textrm{UV}} , 
\\
 \mathcal{U}^{\rm eikonal} & = &   
N_c \frac{1}{\epsilon_\textrm{UV} } .
\end{eqnarray}
\end{subequations}
where $C_F=(N_c^2-1)/(2N_c)$ is the Casimir for the fundamental representation.
In Feynman gauge, the terms  $\mathcal{U}^g$, $\mathcal{U}^Q(s)$,  
$\mathcal{U}^{gQ\bar Q}$, and $\mathcal{U}^{\rm eikonal}$ arise from 
virtual-gluon propagator corrections, virtual-quark propagator corrections, 
quark-gluon vertex corrections, and eikonal line corrections, 
respectively. 
There is one term in Eq.~(\ref{one-loop-div}) with mixed UV and IR poles:
\begin{equation}
\label{eq:calMmixed}
\mathcal{M}(s) = 2N_c \frac{1}{\epsilon_\textrm{UV}\epsilon_\textrm{IR}} 
\left[ 1-\epsilon \log(s/4m^2) \right] .
\end{equation}
In Feynman gauge, this term comes from loop correction to the 
gluon-eikonal vertex.
There are two terms in Eq.~(\ref{one-loop-div}) with only IR poles:
\begin{subequations}
\label{eq:S12IR}
\begin{eqnarray}
\label{eq:S1IR}
\mathcal{S}_1 &=&  2C_F \frac{1}{\epsilon_\textrm{IR}} ,
\\
\label{eq:S2IR}
\mathcal{S}_2(s,z) &=& 2N_c
\left[ -\frac{1}{\epsilon_\textrm{IR}^2}
+ \frac{1}{\epsilon_\textrm{IR}}
 \left( \log(s/4m^2)  + \log(1-z) - \frac{1}{2} \right) \right].
\end{eqnarray}
\end{subequations}
The infrared poles originate from loop-momentum configurations 
in which partons becoming soft and/or collinear. 
The term $\mathcal{S}_1$ is a soft pole that in Feynman gauge
comes from one-loop diagrams obtained from the four LO cut diagrams 
by exchanging a gluon between the on-shell heavy quarks.
All other infrared poles are included in the term $\mathcal{S}_2(s,z)$.

The virtual NLO corrections in Eq.~\eqref{virtual_correction0}
can be expressed as
\begin{eqnarray}
\label{virtual_correction}
 D_{g \to Q \bar Q}^{\rm (virtual)}(z)
 &=& \frac{\alpha_s}{2 \pi}  
 \Gamma(1+\epsilon) \left(\frac{\pi \mu^2}{m^2}\right)^\epsilon 
 \bigg[
\Big( \mathcal{U}^g + 2\mathcal{U}^{gQ\bar Q}
+ \mathcal{U}^{\rm eikonal} + \mathcal{S}_1 \Big)
 D_{g \to Q \bar Q}^{\rm (LO)}(z)
\nonumber \\
&& \hspace{4cm}
+ \int Nd\phi \mathcal{A}_{\textrm{Born}}(p,q)
\Big( \mathcal{U}^Q(s)+ \mathcal{M}(s) + \mathcal{S}_2(s,z) \Big)
\nonumber  \\
&&\hspace{4cm}
+  N_\textrm{CS} \int d\phi_{\rm Born}   
   \mathcal{A}_{\textrm{finite}}(p,q) \bigg],
\end{eqnarray}
where $N d\phi \mathcal{A}_{\rm Born}( p,q)$ is the 
LO differential fragmentation function in Eq.~\eqref{eq:NphiABorn}.
Many of the IR poles cancel against terms in the real NLO corrections,
which are given in Eq.~\eqref{eq:real_emission_subtracted}.
The poles in the real NLO corrections are contained in the 
$D_1$ and $D_2$ subtraction terms in Eq.~\eqref{D1D2_integrated},
the $D_4$ and $D_5$ subtraction terms in Eq.~\eqref{eq:NdPhiD45},
and the $D_3$ subtraction term in Eq.~\eqref{eq:D3int}.
The mixed and subleading poles in $\mathcal{M}(s)$
cancel the 
poles in the $I_2(z)$ term in Eq.~\eqref{eq:Idouble}.
The double IR pole in $\mathcal{S}_2(4m^2,z)$
cancels the $1/\epsilon_\textrm{IR}^2$ poles
in Eq.~\eqref{eq:NdPhiD45}
and in the $\mathcal{V}^{gg}$ term in Eq.~\eqref{def_Vgg}.
The poles $\log(1-z)/\epsilon_\textrm{IR}+\log(s/4m^2)/\epsilon_\textrm{IR}$ 
in $\mathcal{S}_2(4m^2,z)$
also cancel against terms in Eq.~\eqref{eq:NdPhiD45}.
After these cancellations between the real NLO corrections 
and the virtual NLO corrections, the only poles that remain 
are single IR poles proportional to $D_{g \to Q \bar Q}^{\rm (LO)}(z)$
and single UV poles.

\section{Renormalization}
\label{sec:NLOrenorm}

Beyond leading order in $\alpha_s$,
the fragmentation function $D_{g \to \eta_Q}$ depends on a 
factorization scale $\mu_F$ and the running coupling constant $\alpha_s$ 
depends on a renormalization scale $\mu_R$.
If we make those scales explicit, the expansion of the fragmentation function
to NLO in Eq.~\eqref{eq:DNLOeta} becomes
\begin{equation}
\label{eq:DNLO-scales}
D_{g \to \eta_Q}(z,\mu_F) = 
\langle{\cal O}_1(^1S_0)\rangle^{\eta_Q} 
\left[ \alpha_s^2(\mu_R) d_{\rm LO}(z) 
+ \alpha_s^3(\mu_R) d_{\rm NLO}(z,\mu_R,\mu_F) + \ldots \right].
\end{equation}
The scales $\mu_F$ and $\mu_R$ are introduced through renormalization.

The calculation of the fragmentation function is performed in terms of 
the renormalized fields $\Psi_r$ and $A_r$, 
the renormalized coupling constant $g$, 
and the physical mass $m$ of the heavy quark.
Their relations with the corresponding bare quantities 
involve renormalization constants $\delta_2$, 
$\delta_3$, $\delta_g$, and $\delta_m$:
\begin{equation}
\Psi =  (1+\delta_2)^{1/2} \Psi_r, \quad 
A^\mu = (1+\delta_3)^{1/2} A_r^\mu, \quad
g_0 = \mu^\epsilon (1+\delta_g) g, \quad
m_0 = m(1+\delta_m). 
\end{equation}
The renormalization of the  coupling constant is performed in
the $\overline{\textrm{MS}}$ scheme, whereas the renormalization
of the heavy quark mass is performed in the on-shell mass scheme.
In the resulting expressions for the renormalization constants 
$\delta_2$, $\delta_3$, $\delta_g$, and $\delta_m$, it is convenient to 
pull out a common factor:
\begin{equation}
\delta_i = \frac{\alpha_s}{2 \pi} \Gamma(1+\epsilon) \left(\frac{\pi \mu^2}{m^2}  \right)^\epsilon \tilde{\delta}_i .
\end{equation} 
The rescaled renormalization constants $\tilde{\delta}_i$
in the scheme specified above  
read 
\begin{subequations}
\begin{eqnarray}
\tilde{\delta}_2 & = & - \frac{C_F}{2}  \left[\frac{1}{\epsilon_\textrm{UV}}
+\frac{2}{\epsilon_{\textrm{IR}}} + 4 + 6 \log 2  \right] ,
\\  
\tilde{\delta}_3 & = &  
\left(\frac56 N_c - \frac23 T_F n_f \right)
 \left[ \frac{1}{\epsilon_\textrm{UV}}-\frac{1}{\epsilon_\textrm{IR}}  \right] ,
\\ 
\label{eq:deltag}
\tilde{\delta}_g & = & -  \frac{b_0}{2} 
\left[ \frac{1}{\epsilon_\textrm{UV}} + \log \frac{4m^2}{\mu_R^2}  \right] ,
\\
\tilde{\delta}_m &=& - \frac{3 C_F}{2} 
\left[ \frac{1}{\epsilon_\textrm{UV}} + \frac{4}{3}+ 2 \log 2  \right] ,
\end{eqnarray}
\end{subequations}
where  $b_0=(11N_c-4T_F n_f)/6$ 
is the coefficient of $-\alpha_s^2/\pi$ in the beta function 
$(d/d\mu)\alpha_s(\mu)$.
In the counterterm for $g$ in Eq.~\eqref{eq:deltag},
we have allowed for  the renormalization scale $\mu_R$ of $\alpha_s$
to be different from the scale $\mu$ introduced through dimensional regularization.
 
The NLO contributions to the fragmentation function from the 
counterterms for the propagators and vertices in the LO cut diagrams are
\begin{eqnarray}
\label{ct-def}
D_{g \to Q \bar Q}^{\rm (counter)} (z) &=& 
\frac{\alpha_s}{2 \pi} \Gamma(1+\epsilon) \left(\frac{\pi \mu^2}{m^2}  \right)^\epsilon 
 \int N d\phi \mathcal{A}_{\textrm{Born}}(p,q)   
 \left[ 2\mathcal{C}^{gQ\bar Q}+ \mathcal{C}^{\rm{eikonal}} +
 \mathcal{C}^g + \mathcal{C}^Q(s)  \right].
 \nonumber \\
\end{eqnarray}
The terms with the coefficients $2\mathcal{C}^{gQ\bar Q}$, 
$\mathcal{C}^{\rm{eikonal}}$, $\mathcal{C}^g$,
and $\mathcal{C}^Q(s)$ are associated with the quark-gluon vertices, 
the eikonal-gluon vertex, the gluon propagator, and the quark propagator 
in the LO cut diagrams, respectively. The expressions of these coefficients 
in terms of the rescaled renormalization constants $\tilde{\delta}_i$'s are
\begin{subequations}
\label{ct-div}
\begin{eqnarray}
\mathcal{C}^{gQ \bar Q} &=& 2 \tilde{\delta}_g+2\tilde{\delta}_2 + \tilde{\delta}_3 ,
\\  
\mathcal{C}^{\rm{eikonal}} &=& \tilde{\delta}_3, 
\\
\mathcal{C}^g &=& -2\tilde{\delta}_3,
\\
\mathcal{C}^Q(s) &=&  \frac{8m^2}{s-4m^2} \tilde{\delta}_m 
-2  \tilde{\delta}_2 \, .   
\end{eqnarray}
\end{subequations}
The final expression for the counterterm contributions 
to the NLO fragmentation function are  
\begin{eqnarray}
\label{eq:NLOcount}
D_{g \to Q \bar Q}^{\rm (counter)} (z) &=& 
\frac{\alpha_s}{2 \pi} \Gamma(1+\epsilon) \left(\frac{\pi \mu^2}{m^2}  \right)^\epsilon 
\bigg[ \left( 2\mathcal{C}^{gQ\bar Q} + \mathcal{C}^{\rm{eikonal}} + \mathcal{C}^g  \right)
D_{g \to Q \bar Q}^{\rm (LO)}(z)
\nonumber \\
&& \hspace{4cm}
+  \int N d\phi \mathcal{A}_{\textrm{Born}}(p,q)  \mathcal{C}^Q(s) \bigg].
\end{eqnarray}

The field renormalization constants $\tilde{\delta}_2$ and $\tilde{\delta}_3$
include single IR poles.  They cancel the single IR poles that remain after adding 
together the real NLO corrections and the virtual NLO corrections.
The real NLO corrections have single IR poles proportional to 
$D_{g \to Q \bar Q}^{\rm (LO)}(z)$
in the $D_4$ and $D_5$ subtraction terms
in Eq.~\eqref{eq:NdPhiD45}
and in the $\mathcal{V}^{gg}$ and $\mathcal{V}^{q \bar q}$
terms in Eq.~\eqref{def_VggVqq} from the $D_3$ subtraction term.
The virtual NLO corrections have single IR poles proportional to 
$D_{g \to Q \bar Q}^{\rm (LO)}(z)$
in the $\mathcal{S}_1$ and $\mathcal{S}_2(s,z)$ terms in
Eqs.~\eqref{eq:S12IR}.
If the expressions for the coefficients in Eqs.~\eqref{ct-div}
are inserted into the sum of coefficients that appears in Eq.~\eqref{ct-def},
the linear combination of field renormalization constants is
$2 \tilde{\delta}_2 + \tilde{\delta_3}$.
The IR pole in $2  \tilde{\delta}_2$ cancels the IR pole in $\mathcal{S}_1$
introduced in Eq.~(\ref{one-loop-div}). 
The IR pole in $\tilde{\delta}_3$ cancels the yet-to-be-cancelled
single IR poles in the sum of the 
contributions from $ \mathcal{S}_2(s,z)$ introduced in Eq.~(\ref{one-loop-div})
and from the terms $\mathcal{V}^{gg}$ and $\mathcal{V}^{qq}$ in 
Eqs.~(\ref{eq:integralD3}).
This completes the cancellation of the IR poles.

The renormalization of the operator defining the fragmentation function 
also introduces a counterterm.
Its expression in the $\overline{\textrm{MS}}$ scheme reads
\begin{equation}
D_{g \to Q \bar Q}^{\rm (operator)} (z) =
-\frac{\alpha_s}{2 \pi} \Gamma(1+\epsilon)
\left( \frac{\pi \mu^2}{m^2} \right)^\epsilon 
\left[ \frac{1}{\epsilon_\textrm{UV}} + \log \frac{4 m^2}{\mu_F^2}  \right] 
\int_z^1 \frac{dy}{y} P_{gg} (y) 
D_{g \to Q \bar Q}^{\rm (LO)} (z)  ,
\label{operator_ren1}
\end{equation}
where $P_{gg} (y)$ is the Altarelli-Parisi splitting function including both real and virtual contributions:
\begin{equation}
\label{eq:Pgg}
P_{gg}(z)= 2N_c \left[ \frac{z}{(1-z)_+} + \frac{1-z}{z} + z(1-z)  \right]   
+ b_0 \delta(1-z) .
\end{equation}
We have allowed for  the factorization scale $\mu_F$
to be different from the scale $\mu$ introduced through dimensional regularization.

When the contribution from the counterterms 
$D_{g \to Q \bar Q}^{\rm (counter)} (z)$ and $D_{g \to Q \bar Q}^{\rm (operator)} (z)$ 
defined in Eqs.~(\ref{ct-def}) and (\ref{operator_ren1})
are added to the real correction given in Eq.~(\ref{eq:real_emission_subtracted})
and to the virtual correction defined in Eq.~(\ref{virtual_correction}),
all the  poles cancel.
For the UV poles, the cancellation works as follows.
The UV poles originating from the terms  
$\mathcal{U}^g$, $\mathcal{U}^Q(s)$, and $2\mathcal{U}^{gQ \bar Q}$ 
in Eq.~(\ref{one-loop-div}) are canceled by the UV poles from the terms
$\mathcal{C}^g$, $\mathcal{C}^Q(s)$, and  $ 2\mathcal{C}^{gQ \bar Q}$ 
in the expression 
for $D_{g \to Q \bar Q}^{\rm (counter)} (z)$ in Eq.~(\ref{ct-def}).
The poles originating from the term $ I_1(z)$ in Eq.~(\ref{D1D2_integrated})
plus the poles originating from the term  $\mathcal{U}^{\rm{eikonal}}$ 
 in Eq.~(\ref{one-loop-div}) are cancelled by the operator counterterm
$D_{g \to Q \bar Q}^{\rm (operator)} (z)$ in Eq.~(\ref{operator_ren1})
plus the UV poles in the counterterm $\mathcal{C}^{\rm{eikonal}}$ 
introduced in Eq.~(\ref{ct-def}).

\section{Numerical results}

\label{sec:numres}

Once all the poles have been cancelled in the NLO corrections
to the fragmentation function, the function
$d_{\rm NLO}(z,\mu_R,\mu_F)$ in Eq.~\eqref{eq:DNLO-scales}
can be obtained by adding all the finite parts:
\begin{itemize}
\item
the subtracted real NLO corrections,
which are integrated over the phase space 
of the two final-state partons in 4 dimensions,
\item
the finite parts of the integrated subtraction terms
in Eqs.~\eqref{D1D2_integrated}, \eqref{eq:NdPhiD45},
and \eqref{eq:D3int},
\item
the finite parts of the virtual NLO corrections 
in Eq.~\eqref{virtual_correction},
which are integrated over the Born phase space,
\item
the finite parts from the renormalization counterterms in
Eqs.~\eqref{eq:NLOcount} and \eqref{operator_ren1}.
\end{itemize}
The numerical  integrations are performed with the use of the 
 adaptive Monte Carlo integrator Vegas~\cite{Lepage:1980dq}. 

\begin{figure}
\center
\includegraphics[scale=1.2]{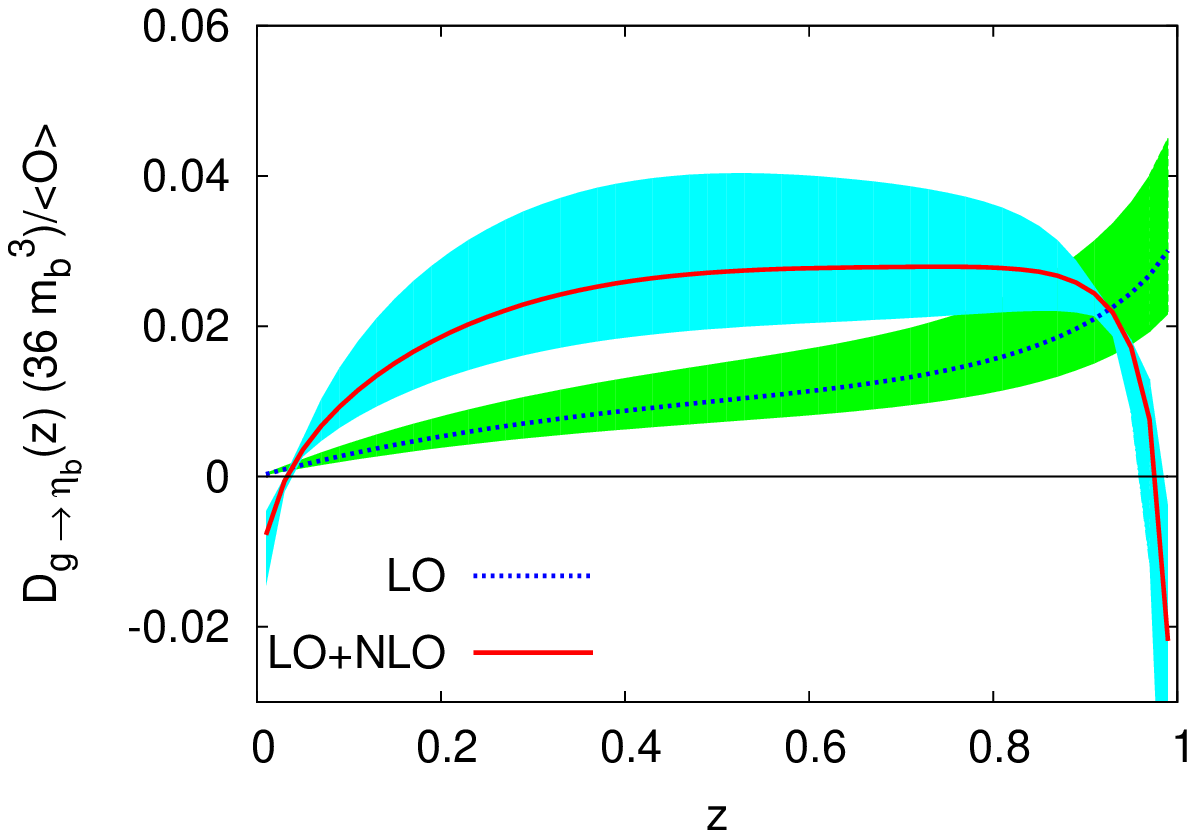}
\caption{ 
The coefficients of $\langle{\cal O}_1(^1S_0)\rangle^{\eta_Q}/(36 m_b^3)$
in the fragmentation function for $g \to \eta_b$ at LO and NLO.
The curves are $\alpha_s^2 d_{\textrm{LO}} (36 m_b^3)$ (dotted line)
and $(\alpha_s^2 d_{\textrm{LO}}+ \alpha_s^3 d_{\textrm{NLO}}) (36 m_b^3)$
(solid line) for the scale choices $\mu_R=\mu_F = 2 m_b$.
The bands are obtained by varying the renormalization scale $\mu_R$ by a factor of 2. }
\label{fig:NumericalResultsR}
\end{figure}

The NLO fragmentation function proportional to
$\alpha_s^2(\mu_R)d_\textrm{LO}(z)
+\alpha_s^3(\mu_R)d_\textrm{NLO}(z,\mu_R,\mu_F)$ 
is compared with the LO fragmentation function
proportional to $\alpha_s^2(\mu_R)d_\textrm{LO}(z)$ 
in Figure~\ref{fig:NumericalResultsR} for the case of bottomonium.
We set  $m_b=4.75$ GeV and $n_f=4$, and we use the value 
$\alpha_s(\mu_R=2m_b)=0.181$ for the strong coupling constant. 
For the central values of the renormalization and factorization scales,
we choose twice the mass of the heavy quark:
$\mu_R=\mu_F = 2 m_b$. 
The LO term $\alpha_s^2d_\textrm{LO}(z)$
increases monotonically from 0 to $\alpha_s^2/(36 m_b^3)$
as $z$ increases from 0 to 1.
The NLO term $\alpha_s^3d_\textrm{NLO}(z)$
increases from $-\infty$ as $z\to 0$
to a broad maximum at an intermediate value of $z$,
and then decreases to $-\infty$ as $z \to 1$.
For $\mu_R=\mu_F = 2 m_b$, its maximum is 
$2.9~\alpha_s^3/(36 m_b^3)$ at $z=0.45$.
The NLO term is negative near both endpoints,
but it cannot be compared to the LO term in these regions, because
the $d_\textrm{NLO}(z)$ is actually a distribution in $z$
with delta-function contributions at the endpoints $z=0$ and $z=1$.
In the integral of the product of $d_\textrm{NLO}(z)$ and a smooth function of $z$,
the endpoint contribution cancels a divergence in the integral up to the endpoint,
so that the integral over the endpoint region is well behaved.
The NLO term can be compared to the LO term in the intermediate region of $z$,
and it is larger than the LO term in this region.
At $z=0.5$, the NLO fragmentation function is larger than the
LO fragmentation function by a factor of 2.7.
The total fragmentation probability obtained by integrating 
the NLO fragmentation function over $z$ from $0$ to $1$
is larger than the LO fragmentation probability by a factor $1.89$ 
for the choice of scales $\mu_R=\mu_F = 2 m_b$.
The mean value $\langle z \rangle$ of the longitudinal momentum fraction
is 2/3 at LO, and it decreases to 0.54 at NLO.

The sensitivity of the LO and NLO fragmentation functions  
to the renormalization scale $\mu_R$ is illustrated in
Figure~\ref{fig:NumericalResultsR}.
The bands are obtained by varying  $\mu_R$ up or down
by a factor of 2 around the central value $2 m_b$
(with $\mu_F=2m_b$). 
The NLO band in Figure~\ref{fig:NumericalResultsR}
is significantly wider than the  LO band
except near the endpoint at $z=1$.
One might have expected the sensitivity to $\mu_R$ 
to be decreased  by adding NLO corrections, but this is not the case
simply because the NLO term in the fragmentation function is
larger than the LO term in the central region of $z$.
The ratio of the  fragmentation functions at NLO and LO is less sensitive 
to the renormalization scale.
At $z=0.5$, the ratio changes from 2.79 to 2.71 to 2.67
as  $\mu_R$ varies from $m_b$ to $2m_b$ to $4m_b$.
The ratio of the fragmentation probabilities at NLO and LO
changes from 1.75 to 1.89 to 1.99.
The mean momentum fraction
$\langle z \rangle$ increases from 0.49 to 0.54 to 0.57.

\begin{figure}
\center
\includegraphics[scale=1.2]{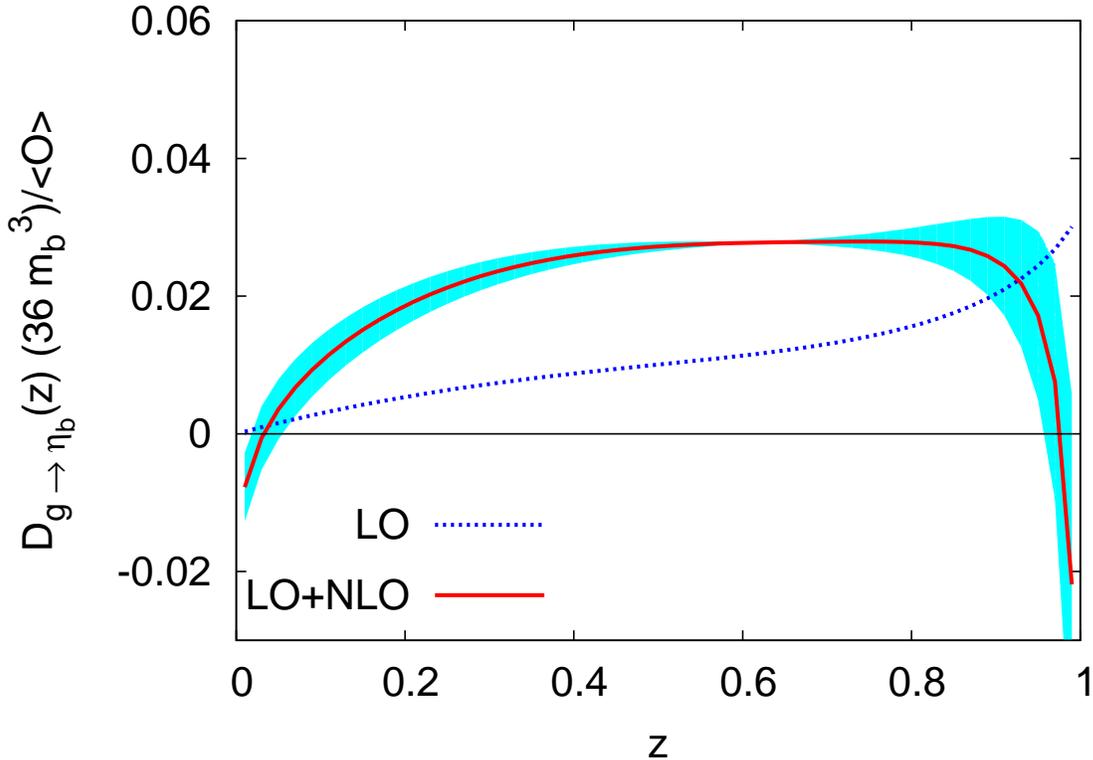}
\caption{ 
The coefficients of $\langle{\cal O}_1(^1S_0)\rangle^{\eta_Q}/(36m_b^3)$
in the fragmentation function for $g \to \eta_b$ at LO and NLO.
The curves are $\alpha_s^2 d_{\textrm{LO}} (36 m_b^3)$ 
(dotted line) and 
$( \alpha_s^2 d_{\textrm{LO}}+ \alpha_s^3 d_{\textrm{NLO}}) (36 m_b^3)$
(solid line) for the scale choices $\mu_R=\mu_F = 2 m_b$.
The band is obtained by varying the factorization scale $\mu_F$ by a factor of 2.}
\label{fig:NumericalResultsF}
\end{figure}

The sensitivity of the NLO fragmentation function 
to the factorization scale $\mu_F$ is illustrated in
Figure~\ref{fig:NumericalResultsF}.
The band is obtained by varying  $\mu_F$ up or down
by a factor of 2 around the central value $2 m_b$
(with $\mu_R=2m_b$). 
In the central region of $z$,
the width of the band from varying $\mu_F$ is much narrower 
than that from varying $\mu_R$ in Figure~\ref{fig:NumericalResultsR}.
The width increases near the endpoints of $z$ at 0 and 1,
but the fragmentation function also has canceling endpoint contributions 
at $z=0$  and $z=1$.
Therefore the increased sensitivity to $\mu_F$ near the endpoints
will not result in a large increase in sensitivity for the integral of 
the product of the fragmentation function and a smooth function of $z$.
The ratio of the fragmentation probabilities at NLO and LO
is more sensitive to $\mu_F$ than to $\mu_R$, ranging from 2.58 to 1.89 to 1.21
as  $\mu_F$ varies from $m_b$ to $2m_b$ to $4m_b$.

\section{Summary}

\label{sec:summary}

In this paper, we have presented the NLO calculation of the fragmentation function 
for a gluon into a spin-singlet S-wave quarkonium state $\eta_Q$ at leading order in $v$. 
This calculation represents 
the first NLO result for a fragmentation function into quarkonium 
that is a nontrivial function of $z$ at LO. 
We have found that the real NLO correction can be organized 
in an efficient way by constructing a set of subtraction terms matching the poles in each 
phase-space boundary leading to singularities. This strategy allows for a transparent 
organization of both UV and IR poles and their cancellation among the different components of the calculation 
(real correction, virtual corrections, and counterterms). It also paves the way to automation 
of the NLO calculation of the fragmentation functions in other NRQCD channels. 
 
We found that the NLO QCD corrections have a dramatic effect on the
fragmentation function in the  $\overline{\textrm{MS}}$ renormalization 
and factorization schemes.  The effect on the shape of the fragmentation function
is particularly dramatic.  Instead of increasing monotonically with $z$ as at LO,
the NLO fragmentation function has a broad maximum in the central region of $z$.
In this region, it is about a factor of 3 larger than at LO.
As a consequence, the NLO fragmentation function displays strong sensitivity 
to the renormalization scale.  
These results suggest that QCD corrections to fragmentation functions could 
have a significant impact on the production of quarkonium states 
at large transverse momentum.   

\begin{acknowledgments}
 
We thank Hong Zhang for useful comments.
P.A. would like to thank Fabio Maltoni for enlightening discussions.
P.A.~is funded in part by the F.R.S.-FNRS Fonds de la Recherche Scientifique (Belgium), 
and by the Belgian Federal Science Policy Office through the Interuniversity Attraction Pole P7/37.
E.B.~was supported in part by the Department of Energy
under grant DE-SC0011726 and by the Simons Foundation.

\end{acknowledgments}

\bibliographystyle{unsrt}
\bibliography{references}

\appendix 

\section{Two-parton phase space integrals}
\label{sec:PhaseSpace}

The real NLO corrections to the fragmentation function
involve integrals over the phase space for two massless partons
whose longitudinal momenta are constrained to add up to $K.n - 2p.n$.  
The dimensionally regularized phase-space measure is
\begin{equation}
\label{eq:dphiq1q2}
d \phi_{\rm real}(p,q_1,q_2)
=\frac{d^{D-1} q_1}{(2 \pi)^{D-1} 2 q_{1,0}} \;
\frac{d^{D-1} q_2}{(2 \pi)^{D-1} 2 q_{2,0}} \;
2 \pi \delta(K.n - (2p + q_1 + q_2).n).
\end{equation}
Explicit parametrizations of this phase-space measure 
are required in 
Sections~\ref{sec:realintUV} and Section~\ref{sec:realintIR}
in order to calculate the poles in $\epsilon$ in the integrals of the
subtraction terms for the real NLO corrections.

The phase-space measure for a single massless parton 
of momentum $q$ can be expressed as
\begin{equation}
\label{eq:dphiq0theta}
\frac{d^{D-1} q}{(2 \pi)^{D-1} 2 q_0} =
\frac{1}{2 (2 \pi)^{3-2\epsilon}} q_0^{1-2\epsilon}dq_0 \;
|\sin\theta|^{1-2\epsilon} d \theta\, d \Omega_\perp ,
\end{equation}
where $\theta$ is the polar angle,
and $d \Omega_\perp$ is the measure 
for integration over angles in the transverse plane.  
The total transverse solid angle is
$\Omega_\perp = 2 \pi^{1-\epsilon}/ \Gamma(1-\epsilon)$.
The light-like 4-vector $n$ defines a spatial direction
that can be used as the polar axis that determines 
the polar angle $\theta$. However 
it is sometimes convenient to choose the polar axis 
in  the spatial direction of a different light-like 4-vector $k$.
A convenient alternative set of variables 
from $q_0$ and $\theta$ is the longitudinal momentum
$q.n$ and a variable $\lambda$ defined by
\begin{equation}
\label{eq:lambda}
\lambda = 2k.q/k.n .
\end{equation}
The phase-space measure can be expressed as
\begin{equation}
\label{eq:dphiqnlambda}
\frac{d^{D-1} q}{(2 \pi)^{D-1} 2 q_0} =
\frac{1}{4 (2 \pi)^{3-2\epsilon}}\, (q.n)^{- \epsilon} d(q.n)\, 
\lambda^{-\epsilon} d \lambda\;d \Omega_\perp .
\end{equation}
This phase-space measure is independent of the overall scales  
of $n$ and $k$.
In a Lorentz frame where the spatial parts of $n$ and $k$ 
are back-to-back, $q.n = q_0n_0(1 - \cos\theta)$ 
and $\lambda = q_0(1  + \cos\theta)/n_0$,
so Eq.~\eqref{eq:dphiqnlambda} reduces to the measure in
Eq.~\eqref{eq:dphiq0theta}.

A convenient  parameterization of the two-parton phase-space measure in 
Eq.~\eqref{eq:dphiq1q2} can be obtained by introducing
two light-like 4-vectors that define the polar axes of $q_1$ and $q_2$:
a 4-vector $k_1$ that may depend on $p$ and 
a 4-vector $k_2$ that may depend on $p$  and $q_1$.
We also introduce a momentum fraction $u$ defined by
\begin{equation}
\label{eq:IRu}
u = \frac{q_2.n}{(q_1+q_2).n} .
\end{equation}
After integrating over the longitudinal component and transverse angles of $q_1$,
the phase-space measure reduces to 
\begin{equation}
\label{eq:dphiulambda}
d \phi_{\rm real}(p,q_1,q_2) =
\frac{2^{-2\epsilon}[(1-z)K.n]^{1-2\epsilon}}{(4 \pi)^{4-3 \epsilon}\Gamma(1-\epsilon)}\, 
[u(1-u)]^{-\epsilon} du\; \lambda_1^{-\epsilon} d \lambda_1\; 
\lambda_2^{-\epsilon} d \lambda_2\; d \Omega_{2\perp} .
\end{equation}

{\bf Phase space for UV and mixed poles.}
To obtain the phase-space parameterization used to integrate 
the UV and mixed poles in Section~\ref{sec:realintUV},
we choose the light-like vectors $k_1$ and $k_2$ that specify the polar axes 
for $q_1$ and $q_2$ to be
\begin{equation}
\label{eq:UVk1k2}
k_1^\mu = 2p^\mu - \frac{m^2}{p.n} n^\mu, \qquad
k_2^\mu = (2p+q_1)^\mu - \frac{(2p+q_1)^2}{2(2p+q).n} n^\mu.
\end{equation}
The variables $\lambda_1$ and $\lambda_2$ defined by 
Eq.~\eqref{eq:lambda} are
\begin{eqnarray}
\lambda_1 & = & \frac{1}{zK.n}  
\left(s_1  - \frac{1-u+uz}{z}4m^2 \right) ,
\\  
\lambda_2 & = &  
\frac{1}{(1-u+uz)K.n}   
 \left( s -  \frac{1}{1-u+uz} s_1 \right)  ,
\end{eqnarray}
where $s_1 = (2p+q_1)^2$ and $s = (2p+q_1+q_2)^2$.
We then change variables in Eq.~\eqref{eq:dphiulambda}
from $u$, $\lambda_1$, and $\lambda_2$ 
to $y_1 = 1-u+uz$, $s_1$, and $s$.
The phase-space measure reduces to
\begin{eqnarray}
\label{eq:dphiulambdaUV}
d \phi_{\rm real}(p,q_1,q_2) &=&
\frac{2^{-2\epsilon} z^{-1+\epsilon}}{(4 \pi)^{4-3 \epsilon}\Gamma(1-\epsilon)K.n}\, 
y_1^{-1+\epsilon} (1-y_1)^{-\epsilon} (y_1-z)^{-\epsilon} dy_1\; (s-s_1/y_1)^{-\epsilon} d s\; 
\nonumber \\
&& \hspace{4cm}
 \times [s_1 - 4m^2/(z/y_1)]^{-\epsilon} d s_1\; d \Omega_{2\perp} .
\end{eqnarray}
After multiplying by the Collins-Soper prefactor in 
Eq.~\eqref{eq:overalfac}, we obtain the measure given by
Eqs.~\eqref{eq:Ndphi_UV} and \eqref{eq:def_dphi1_UV} with $i=1$.

{\bf Phase space for IR poles.}
To obtain the phase-space parameterization used to integrate 
the IR poles in Section~\ref{sec:realintIR}, 
we first introduce additional integrals over a light-like 4-vector $\tilde q$ that has the same
longitudinal component as $q_1+q_2$ and over the invariant mass 
$(q_1+q_2)^2$.  We do this by multiplying the 
phase-space measure in Eq.~\eqref{eq:dphiq1q2} by 1
in the form
\begin{equation}
\label{eq:tildeqint}
1 = \int_0^\infty dt \int d^D \tilde{q}\, 
\delta^D(\tilde q- (q_1+q_2) + [t/2(q_1+q_2).n]n)\;
\delta(t - (q_1+q_2)^2) .
\end{equation}
After expressing the phase-space measure for $q_1$ in the 
manifestly covariant form 
$d^D q_1 \delta(q_1^2) \theta(q_1.n)/(2 \pi)^{D-1}$,
the $D$-dimensional delta function  in Eq.~\eqref{eq:tildeqint} 
can be used to integrate over $q_1$.
The phase-space measure in Eq.~\eqref{eq:dphiq1q2} 
can then be reduced to
\begin{equation}
\label{eq:dphiq1q2IR}
d \phi_{\rm real}(p,q_1,q_2)
=\frac{d^{D-1} \tilde q}{(2 \pi)^{D-1} 2 \tilde q_0} \;
\frac{d^{D-1} q_2}{(2 \pi)^{D-1} 2 q_{2,0}} \;
2 \pi \delta(K.n - (2p + \tilde q).n) \frac{\tilde q.n}{(\tilde q - q_2).n}.
\end{equation}
The last factor comes from integrating $ \delta(q_1^2)$ over $t$
and is equal to $1/(1-u)$, where $u = q_2.n/\tilde q.n$.
We choose the light-like vectors $\tilde k$ and $k_2$
that specify the polar axes for $\tilde q$ and $q_2$ to be
\begin{equation}
\label{eq:IRk1k2}
\tilde k^\mu = 2p^\mu - \frac{m^2}{p.n} n^\mu, \qquad
k_2^\mu = \tilde q^\mu.
\end{equation}
The variables $\tilde \lambda$ and $\lambda_2$ defined by 
Eq.~\eqref{eq:lambda} are
\begin{eqnarray}
\tilde \lambda & = & \frac{1}{zK.n}  
\left(\tilde s  - \frac{1}{z}4m^2 \right) ,
\\  
\lambda_2 & = &  \frac{(1-u)t}{(1-z)K.n} ,
\end{eqnarray}
where $\tilde s = (2p+\tilde q)^2$ and $t = 2\tilde q.q_2/(1-u)$.
We insert the expressions analogous to Eq.~\eqref{eq:dphiulambda} 
for the phase-space measures of $\tilde q$ and $q_2$
into Eq.~\eqref{eq:dphiq1q2IR}.
After integrating over the longitudinal component and transverse angles of $\tilde q$,
the measure becomes
\begin{equation}
\label{eq:dphiulambdaIR1}
d \phi_{\rm real}(p,q_1,q_2) =
\frac{2^{-2\epsilon}[(1-z)K.n]^{1-2\epsilon}}{(4 \pi)^{4-3 \epsilon}\Gamma(1-\epsilon)}\, 
\frac{u^{-\epsilon}}{1-u} du\; \tilde \lambda^{-\epsilon} d \tilde\lambda\; 
\lambda_2^{-\epsilon} d \lambda_2\; d \Omega_{2\perp} .
\end{equation}
We then change variables from $\tilde\lambda$ and $\lambda_2$ 
to $\tilde s$ and $t$.  The phase-space measure reduces to
\begin{eqnarray}
\label{eq:dphiulambdaIR2}
d \phi_{\rm real}(p,q_1,q_2) =
\frac{2^{-2\epsilon} z^{-1+\epsilon}(1-z)^{-\epsilon}}
       {(4 \pi)^{4-3 \epsilon}\Gamma(1-\epsilon)K.n}\, 
[u(1-u)]^{-\epsilon} du\; (\tilde s - 4m^2/z)^{-\epsilon} d \tilde s\; 
t^{-\epsilon} d t\; d \Omega_{2\perp} .
\end{eqnarray}
After multiplying by the Collins-Soper prefactor in 
Eq.~\eqref{eq:overalfac} and making the change of variables $\lambda = t/4m^2$, 
we obtain the measure given by
Eqs.~\eqref{eq:Ndphi_IR} and \eqref{eq:dtildephi_IR}.

\section{Virtual loop integrals with an eikonal propagator}
\label{sec:LoopIntegrals}

The virtual NLO corrections to the fragmentation function for $g \to Q \bar Q$
require the evaluation of loop integrals whose integrand is the product 
of Feynman propagators and a single eikonal propagator.
These integrals have UV and IR divergences, which in dimensional regularization
appear as poles in $\epsilon = (4-D)/2$.
In this Appendix, we present the pole terms in these integrals.

\subsection{Reduction to Feynman parameter integrals}

The single eikonal propagator can be expressed in the form
$1/[(l+P).n + i \varepsilon]$, where $l$ is the loop momentum 
and the 4-vector $P$ is a linear combination of $p$ and $q$.
After combining the $j$ Feynman denominators using Feynman parameters,
the loop integral can be expressed in the form
\begin{equation}
\label{eq:loopA}
\int \frac{d^D l}{(2 \pi)^D}
\frac{1}{[(l - Q)^2 - \Delta + i \varepsilon]^j} 
\frac{1}{(l+P).n + i \varepsilon},
\end{equation}
where the 4-vector $Q$  is a linear combination of $p$ and $q$ 
whose coefficients depend on Feynman parameters,
and the scalar  $\Delta$ is a linear combination of $m^2$ and $p.q$ 
whose coefficients depend on Feynman parameters.
The eikonal denominator can be combined with the other denominator
by introducing an additional integral over a variable $\lambda$:
\begin{equation}
\label{eq:loopB}
\int \frac{d^D l}{(2 \pi)^D} (2j) \int_0^\infty d\lambda
\frac{1}{[(l - Q)^2 + 2 \lambda (l+P).n - \Delta + i \varepsilon]^{j+1}}.
\end{equation}

After making the shift $l \to l+Q-\lambda n$ in the loop momentum
and then  evaluating the integral over $\lambda$, the result is
\begin{equation}
\label{eq:loopC}
\frac{1}{(Q+P).n + i \varepsilon}
\int \frac{d^D l}{(2 \pi)^D}
\frac{1}{[l^2 - \Delta + i \varepsilon]^j} .
\end{equation}
The eikonal denominator in Eq.~\eqref{eq:loopA}
has been replaced by one that is independent of the loop momentum $l$
but depends on the Feynman parameters.
After evaluating the integral over the loop momentum, the result is
\begin{equation}
\label{eq:loopD}
\frac{(-1)^j i }{(4\pi)^{D/2}} \frac{\Gamma(j-D/2)}{\Gamma(j)}
(\Delta - i \varepsilon)^{D/2-j} 
\frac{1}{(Q+P).n + i \varepsilon}.
\end{equation}
It remains only to evaluate the integrals over the Feynman parameters.

If the loop integral is multiplied by $(2p+q).n$,
it is independent of the scale of the light-like vector $n$.
Since it is Lorentz invariant function of $p$, $q$, and $n$, 
the resulting integral must be $(m^2)^{D/2-j}$ 
multiplied by a function of two dimensionless variables:
\begin{equation}
\label{eq:zrdef}
z = \frac{(2p).n}{(2p+q).n}, \qquad r = \frac{p.q}{m^2}.
\end{equation}
If $j=2$, the integral over the loop momentum is UV divergent,
resulting in the factor $\Gamma(\epsilon)$ in Eq.~\eqref{eq:loopD}.
The integrals over the Feynman parameters may also give IR divergences.
These divergences can be isolated into terms 
that can be evaluated analytically, giving poles in $\epsilon$.
The finite terms of order $\epsilon^0$ can also be evaluated analytically 
in terms of logarithms and dilogarithms of functions of $z$ and $r$.
For some of the loop integrals, the evaluation of the integrals using a 
computer algebra program, such as Mathematica,
gives dilogarithms with many different arguments.
The number of different arguments can be reduced 
by using functional identities for dilogarithms.
However, for many of the integrals, the expressions for the finite terms 
are still sufficiently complicated that we do not present them here.

\subsection{Integrals with two Feynman propagators}

There are 12 independent integrals with two Feynman propagators 
and an eikonal propagator.
These integrals have a UV divergence that yields a single pole in $\epsilon$.
They may also have an IR divergence that yields a second pole in $\epsilon$.
If there is a double pole in $\epsilon$, 
the nature of the subleading single pole is ambiguous.
It is convenient to express these integrals as the product of a Laurent expansion 
in $\epsilon$ and  an overall factor of
\begin{equation}
I_2 = i \frac{(4\pi)^{\epsilon}\Gamma(1+\epsilon)}{16 \pi^2 (2p+q).n} 
(4 m^2)^{-\epsilon}.
\end{equation}
It is convenient to introduce a compact notation for the Feynman propagators.
The propagators for momentum $k$ and masses 0 and $m$ are denoted by
\begin{subequations}
\begin{eqnarray}
D(k) &=& \frac{1}{k^2 + i \varepsilon},
\\
D(k,m) &=& \frac{1}{k^2 - m^2 + i \varepsilon}.
\end{eqnarray}
\end{subequations}

The pole terms in the 10 nonzero integrals are
\begin{subequations}
\begin{eqnarray}
\int\frac{d^D l}{(2\pi)^D}  \frac{D(l) D(l-2p)}{l.n+i\varepsilon} 
&=&   \frac{I_2}{z} \left[-  \frac{1}{\epsilon_{\textrm{UV}}\epsilon_{\textrm{IR}}} 
+ \frac{1}{\epsilon}(-i\pi)
+ {\cal O}(\epsilon^0) \right],
\\
\int\frac{d^D l}{(2\pi)^D}   \frac{D(l) D(l-2p)}{(l+q).n+i\varepsilon} 
&=&   \frac{I_2}{z} \left[- \frac{1}{\epsilon_{\textrm{UV}}} \log (1-z)
+ {\cal O}(\epsilon^0)  \right],
\\
\int\frac{d^D l}{(2\pi)^D}  \frac{D(l) D(l-p,m)}{l.n+i\varepsilon} 
&=&   \frac{I_2}{z} \left[-  \frac{1}{\epsilon_{\textrm{UV}}\epsilon_{\textrm{IR}}}  
-  \frac{2}{\epsilon} \log 2+ {\cal O}(\epsilon^0) \right],
\\
\int\frac{d^D l}{(2\pi)^D}  \frac{D(l) D(l-p,m)}{(2p-l).n+i\varepsilon}
&=&   \frac{I_2}{z}  \left[ \frac{2}{\epsilon_{\textrm{UV}}} \log 2 
+ {\cal O}(\epsilon^0)  \right],
\\
\int\frac{d^D l}{(2\pi)^D}  \frac{D(l) D(l-p,m)}{(l+q).n+i\varepsilon} 
&=&  \frac{I_2}{z}  \left[ \frac{2}{\epsilon_{\textrm{UV}}} \log \frac{2-z}{2(1-z)}
+ {\cal O}(\epsilon^0)  \right],
\\
\int\frac{d^D l}{(2\pi)^D}  \frac{D(l) D(l-p,m)}{(2p+q-l).n+i\varepsilon} 
&=&   \frac{I_2}{z} \left[ - \frac{2}{\epsilon_{\textrm{UV}}} \log \frac{2-z}{2} 
+ {\cal O}(\epsilon^0)  \right],
\\
\int\frac{d^D l}{(2\pi)^D}  \frac{D(l) D(l-2p-q)}{l.n+i\varepsilon}
 &=&  I_2
\left[-  \frac{1}{\epsilon_{\textrm{UV}}\epsilon_{\textrm{IR}}} 
+ \frac{1}{\epsilon} [ \log(1+r)  - i\pi]
+ {\cal O}(\epsilon^0)  \right],
\\
\int\frac{d^D l}{(2\pi)^D} \frac{D(l) D(l-p-q,m)}{l.n+i\varepsilon} 
&=&   \frac{I_2}{2-z}
\left[ - \frac{2}{\epsilon_{\textrm{UV}}\epsilon_{\textrm{IR}}} 
+ \frac{2}{\epsilon} \left( \log \frac{r}{2} - i \pi   \right)
+ {\cal O}(\epsilon^0)  \right],
\\
 \int\frac{d^D l}{(2\pi)^D} \frac{D(l) D(l-p-q,m)}{(2p+q-l).n+i\varepsilon}
&=&   \frac{I_2}{2-z}
\left[  - \frac{2}{\epsilon_{\textrm{UV}}} \log \frac{z}{2}
+ {\cal O}(\epsilon^0) \right],
\\
\int\frac{d^D l}{(2\pi)^D} \frac{D(l,m) D(l-q,m)}{(l+p).n+i\varepsilon} 
&=&   \frac{I_2}{1-z}
\left[  \frac{1}{\epsilon_{\textrm{UV}}} \log \frac{2-z}{z}  
+ {\cal O}(\epsilon^0) \right].
\end{eqnarray}
\end{subequations}
There are also two integrals that vanish with dimensional regularization,
because the loop integral in Eq.~\eqref{eq:loopC} has no scale.
The vanishing of these integrals can be interpreted as due to
cancellations between poles in $\epsilon$ that are of UV and IR origin:
\begin{subequations}
\begin{eqnarray}
\int\frac{d^D l}{(2\pi)^D}  \frac{D(l) D(l-q)} {l.n+i\varepsilon}
&=&  \frac{ I_2}{1-z} 
\left[- \left( \frac{1}{\epsilon_{\textrm{UV}} \epsilon_{\textrm{IR}}} 
-  \frac{1}{\epsilon_{\textrm{IR}}^2} \right) \right],
\\
\int\frac{d^D l}{(2\pi)^D}  \frac{D(l) D(l-q)}{(l+2p).n+i\varepsilon} 
 &=&  \frac{ I_2}{1-z}
  \left[ -\left( \frac{1}{\epsilon_{\textrm{UV}}} 
  -  \frac{1}{\epsilon_{\textrm{IR}}}  \right)  \log z \right].
\end{eqnarray}
\end{subequations}

\subsection{Integrals with three Feynman propagators}

There are 8 independent integrals with three Feynman propagators 
and an eikonal propagator.
These integral have no UV divergences.
The IR divergences yield double and single poles in $\epsilon$.
It is convenient to express these integrals as the product of a Laurent expansion 
in $\epsilon$ and  an overall factor of
\begin{equation}
I_3= i \frac{(4\pi)^{\epsilon}\Gamma(1+\epsilon)}{16 \pi^2 (2p+q).n}
(4m^2)^{-1-\epsilon}.
\end{equation}

The pole terms in the 8 independent integrals are
\begin{subequations}
\begin{eqnarray}
&&\int \frac{d^Dl}{(2\pi)^{D}} 
\frac{D(l) D(l-p,m) D(l-p-q,m)}{l.n+i\varepsilon} 
\nonumber \\ 
&& \hspace{1.5cm}
=  \frac{I_3}{zr} \left[ 
\frac{2}{\epsilon^2_{\textrm{IR}}} +\frac{4}{\epsilon_{\textrm{IR}}} 
\left( -\log r + \log\frac{2-z}{z}  +i \pi    \right)    
+ {\cal O}(\epsilon^0) \right],
\\
& & 
\int \frac{d^Dl}{(2\pi)^{D}} 
\frac{D(l) D(l-p,m) D(l-2p-q,m)}{l.n+i\varepsilon}\nonumber \\
&& \hspace{1.5cm}
=  \frac{I_3}{z(r+1)} \left[
\frac{1}{\epsilon^2_{\textrm{IR}}} +\frac{2}{\epsilon_{\textrm{IR}}} 
\left( -\log (r+1) - \log z + i \pi    \right) 
+ {\cal O}(\epsilon^0) \right],
\\
& & 
\int \frac{d^Dl}{(2\pi)^{D}} 
\frac{D(l) D(l-2p-q) D(l-p-q,m)}{l.n+i\varepsilon} 
\nonumber \\
&& \hspace{1.5cm}
=  \frac{I_3}{(1-z)r+2-z} \left[
 \frac{2}{\epsilon_{\textrm{IR}}} 
\left( -\log\frac{r+1}{r} - \log(2-z)  \right)
+ {\cal O}(\epsilon^0) \right],
\\
& & 
\int \frac{d^Dl}{(2\pi)^{D}} 
\frac{D(l) D(l-q) D(l-2p-q)}{l.n+i\varepsilon} 
\nonumber \\
&& \hspace{1.5cm}
=  \frac{I_3}{(1-z)(r+1)} \left[
\frac{1}{\epsilon^2_{\textrm{IR}}} 
+ \frac{1}{\epsilon_{\textrm{IR}}} 
\left( -2 \log (r+1) - \log(1-z) + i \pi   \right)
+ {\cal O}(\epsilon^0) \right],
\nonumber \\
\\
& &
\int \frac{d^Dl}{(2\pi)^{D}} 
\frac{D(l) D(l-p,m) D(l-p-q,m)}{(2p+q-l).n+i\varepsilon} 
= {\cal O}(\epsilon^0),
\\
& &
 \int \frac{d^Dl}{(2\pi)^{D}} 
\frac{D(l) D(l-q) D(l-p-q,m)}{l.n+i\varepsilon} 
\nonumber \\
&& \hspace{1.5cm}
=  \frac{I_3}{(1-z)r} \left[
 \frac{3}{\epsilon_\textrm{IR}^2}
+\frac{2}{\epsilon_\textrm{IR}} 
\left( -2 \log r+ \log\frac{2-z}{1-z} + 2 i \pi \right)
+ {\cal O}(\epsilon^0) \right],
\\
& &
 \int \frac{d^Dl}{(2\pi)^{D}} 
\frac{D(l) D(l-q) D(l-2p-q)}{(2p+q-l).n+i\varepsilon} 
\nonumber \\
&& \hspace{1.5cm}
=  \frac{I_3}{zr -1+z} \left[
- \frac{2}{\epsilon_\textrm{IR}} \left( \log (r+1) + \log z \right)
+ {\cal O}(\epsilon^0) \right],
\\
& &
 \int \frac{d^Dl}{(2\pi)^{D}} 
\frac{D(l) D(l-q) D(l-p-q,m)}{(2p+q-l).n+i\varepsilon} 
\nonumber \\
&& \hspace{1.5cm}
=  \frac{I_3}{zr} \left[
\frac{1}{\epsilon_{\textrm{IR}}^2} 
+ \frac{2}{\epsilon_{\textrm{IR}}} \left( - \log r - \log z  + i \pi \right) 
+ {\cal O}(\epsilon^0) \right].
\end{eqnarray}
\end{subequations}

\end{document}